\documentclass[twocolumn, twocolappendix]{aastex701}

\usepackage{xspace}
\usepackage{amsmath}
\usepackage{multirow}
\usepackage{txfonts}
\newcommand{\revise}[1]{\textrm{#1}}
\newcommand{\rrevise}[1]{\textrm{#1}}

\shorttitle{NH$_3$ in IRAS 16293}
\shortauthors{Yamato et al.}
\graphicspath{{./}}

\begin{document}

\title{Subarcsecond Multi-line Observations of NH$_3$ with VLA toward the Class 0 Source IRAS~16293-2422}

\author[0000-0003-4099-6941]{Yoshihide Yamato}
\affiliation{Department of Astronomy, Graduate School of Science, The University of Tokyo, 7-3-1 Hongo, Bunkyo, Tokyo 113-0033, Japan}
\affiliation{RIKEN Pioneering Research Institute, 2-1 Hirosawa, Wako, Saitama 351-0198, Japan}
\email[show]{yyamato.as@gmail.com}

\author[0000-0003-3283-6884]{Yuri Aikawa}
\affiliation{Department of Astronomy, Graduate School of Science, The University of Tokyo, 7-3-1 Hongo, Bunkyo, Tokyo 113-0033, Japan}
\email{}

\author[0000-0002-2026-8157]{Kenji Furuya}
\affiliation{Department of Astronomy, Graduate School of Science, The University of Tokyo, 7-3-1 Hongo, Bunkyo, Tokyo 113-0033, Japan}
\affiliation{RIKEN Pioneering Research Institute, 2-1 Hirosawa, Wako, Saitama 351-0198, Japan}
\email{}

\author[0000-0002-6195-0152]{John J. Tobin}
\affiliation{National Radio Astronomy Observatory, 520 Edgemont Road, Charlottesville, VA 22903, USA}
\email{}

\author[0000-0001-9133-8047]{Jes K. J{\o}rgensen}
\affiliation{Niels Bohr Institute, University of Copenhagen, Øster Voldgade 5-7, 1350, Copenhagen, Denmark}
\email{}

\begin{abstract}
Ammonia (NH$_3$) is one of the key volatiles that plays a central role in nitrogen chemistry and its evolution during the epoch of star and planet formation. We present subarcsecond ($\sim0\farcs5$) resolution observations of NH$_3$ molecular emission lines with Karl G. Jansky Very Large Array (VLA) toward the Class 0 multiple system IRAS 16293-2422 including source A and source B as major components. This comprises the most comprehensive set of NH$_3$ line observations in protostellar sources to date, which includes 17 inversion transitions with a wide range of upper state energies ($E_\mathrm{u}$) spanning from $\sim$23\,K to $\sim$1,580\,K. We detect spatially resolved emission of a number of transitions, 
and find that the high-$E_\mathrm{u}$ ($\gtrsim$1,000\,K) lines show compact distributions in the vicinity of protostars while low-$E_\mathrm{u}$ ($\lesssim$150\,K) lines exhibit more extended emission. Utilizing a two-component model, we constrain the rotation temperature and NH$_3$ column density for both source A and source B. The rotation temperature of the warmer component reaches $\sim200$--300\,K, indicating that the high-$E_\mathrm{u}$ lines selectively trace the inner hot region. We suggest that this hot NH$_3$ gas in source A is originated from the local shock heating based on the comparison with the previous high-resolution ALMA observations, while that in source B could be explained by the mass accretion heating in the innermost hot region. We also briefly discuss the chemistry related to NH$_3$ based on the abundance ratios relative to major icy molecules derived using literature values.


\end{abstract}



\section{Introduction}\label{sec:intro}
The physical and chemical structures around low-mass protostars are crucial in determining the initial condition of the planetary system formation. Recent spatially resolved observations with Atacama Large Millimeter/submillimeter Array (ALMA) have mapped the distributions of various molecular gas from the envelope scale to the protostellar disk scale in detail, revealing the physical components each molecular line tends to trace \citep[e.g.,][]{Ohashi2014, Sakai2014, Oya2018, Tychoniec2021, Sharma2025}. CO (and its isotopologues) is the most frequently observed molecule and the major reservoir of carbon and oxygen, which traces the various components including outflows, envelopes, and protostellar disks \citep[e.g.,][]{Yen2014, Aso2015}. Sulfur-bearing molecules, SO, have been proposed as a weak shock tracer, which are detected at the disk-envelope interface where the envelope materials accrete onto the disk \revise{\citep[e.g.,][]{Sakai2014_Nature, Ohashi2014, vanGelder2021, Liu2025}}. Complex organic molecules (COMs), mainly oxygen-bearing ones, are now frequently detected in the inner warm envelope by \revise{(sub-)mm interferometric observations \citep[e.g.,][]{Jorgensen2016, Lopez-Sepulcre2017, Bianchi2020, Belloche2020, Yang2021, Martin-Domenech2021, Bouvier2022, Hsu2026}}


However, nitrogen-bearing species are less focused on, despite \rrevise{their} importance for the chemical evolution of the interstellar medium (ISM) and the prebiotic chemistry \citep{Daranlot2012}. In particular, we critically lack the observations of the major nitrogen reservoir. 
\revise{Although a growing number of nitrogen-bearing molecules and even larger nitrogen-bearing COMs, such as NH$_2$CHO, CH$_3$CN and CH$_3$CH$_2$CN, have been detected toward a handful of sources \citep[e.g.,][]{Tychoniec2021, Nazari2021, Martin-Domenech2021},}
they only account for a small portion of nitrogen in the ISM. The major nitrogen reservoir in the ISM could be either atomic nitrogen N or molecular nitrogen N$_2$, but direct observations of them is challenging as they do not emit efficiently in the cold environment. 

Ammonia (NH$_3$) is another primary nitrogen reservoir in the ISM, which accounts for up to $\sim10$\% of the overall nitrogen in molecular clouds \citep{Oberg2011}, and a number of inversion transitions are available at cm wavelengths. The bulk of NH$_3$ forms on the ice mantles in the early stages of star formation, and will be delivered to the protostellar environment \citep[e.g.,][]{Furuya2018}. Given its similar volatility to water and COMs, NH$_3$ is expected to sublimate in the inner ($\lesssim100$\,au) warm ($\gtrsim100$\,K) region of the protostellar envelopes. Indeed, using the Karl G. Jansky Very Large Array (VLA), \citet{Choi2010} detected \revise{marginally-spatially-resolved,} compact \revise{($\sim$100\,au)} emission of NH$_3$ (2,2) and (3,3) inversion transitions at $\sim0\farcs3$ (or $\sim100$\,au) resolution toward the protobinary system NGC~1333 IRAS~4A \citep[see also][]{Choi2007}. \citet{Yamato2022} observed several NH$_3$ (and its deuterated isotopologue NH$_2$D) inversion transitions at $\sim1\arcsec$ (or $\sim300$\,au) resolution toward the same source, which enabled \revise{a} multi-line excitation analysis. The excitation temperature of NH$_3$ lines are derived to be $\gtrsim100$\,K, consistent with that of CH$_3$OH \citep{DeSimone2020, DeSimone2022}. While these observations likely trace the warm NH$_3$ gas in the vicinity of the protostars, spatially-resolved, multi-line observations are needed to better constrain the precise distributions and excitation conditions of NH$_3$.



In this paper, we present the first high-resolution, multi-line NH$_3$ observations with VLA toward the Class 0 source IRAS 16293-2422. IRAS 16293-2422 is one of the most well-studied low-mass protostellar systems located in the Ophiucus star-forming region ($d\sim140$\,pc; \citealt{Dzib2018}). This system consists of IRAS 16293-2422 A (hereafter source A) and IRAS 16293-2422 B (hereafter source B), separated by $\approx5\arcsec$ (or $\sim700$\,au) on the plane of sky, where source A itself has been revealed to be the bounded binary A1 and A2 \citep{Maureira2020}. \revise{While source A harbors circumbinary/circumstellar disks with individual stellar masses of $\sim$0.5--2.0\,$M_\odot$ for A1 and A2 \citep{Maureira2020}, source B lacks the detection of \rrevise{a} clear disk-like rotation signature (likely due to its face-on geometry) and thus \rrevise{has a} poorer stellar mass estimate ($\sim$0.1--1\,$M_\odot$; e.g., \citealt[][]{Oya2018}).}
The distributions of the molecular gas around the protostars have been extensively studied with high-resolution interferometric observations at millimeter/submillimeter wavelengths \citep{Jacobsen2018, Oya2016, Oya2018, Oya2020}, including the ALMA Protostellar Interferometric Line Survey (PILS; \citealt{Jorgensen2016}) that have detected numerous lines of COMs in the warm inner envelope. IRAS 16293-2422 has also been observed frequently in dust continuum emission with VLA \citep{Loinard2002, Chandler2005, Loinard2007, Pech2010, Hernandez-Gomez2019}, revealing its complex structure particularly in source A. The first interferometric NH$_3$ line observations toward this source have been performed in ($J, K$) = (1,1) and (2,2) transitions with VLA at 6\arcsec--13\arcsec resolution \citep{Mundy1990}, tracing the extended, cold (15--20\,K) component in the outer envelope at $\gtrsim1000$\,au scale. No subarcsecond resolution observations of NH$_3$ lines, which can probe the warm inner envelope, have been reported to date.   

This paper is structured as follows. Section \ref{sec:observation} describes the details of observations and data reduction. The observational results and the multi-line analysis of the NH$_3$ lines are presented in Section\revise{s} \ref{sec:results} and \ref{sec:analysis}, respectively. We discuss the results in Section \ref{sec:discussion} and finally summarize the study in Section \ref{sec:summary}.


\section{Observations and Data reduction} \label{sec:observation}
We observed the IRAS 16293-2422 system with \revise{the} VLA during the Semester 2022B (PI: Y. Yamato, project ID: 22B-219). Two observations were performed on 9 and 12 January 2023 during the transition from antenna configuration C to B with the majority of antennas in their B-configuration positions on the north and east arms. The baseline coverage was from 102\,m to 11.1\,km. While 27 antennas were used for the first track, there were 26 antennas operating for the second track. The duration of the executions were 1.5 hours for both tracks, which resulted in a total on-source time of 102 min. For both execution blocks, quasar 3C286 was used as the flux and bandpass calibrator while the phase and amplitude calibrator was quasar J1625-2527. The 3-bit correlator setup in the K band receiver was used and included 63 spectral windows, 31 of which were used to target spectral lines with a narrow bandwidth (8 or 16 MHz) and a fine channel width (7.8, 15.6, or 31.3 kHz\revise{; corresponding to $\approx$0.1--0.5 km\,s$^{-1}$}). A total of 17 NH$_3$ transitions were covered by these windows for spectral lines. These NH$_3$ transitions have a wide range of excitation properties (upper state energy $E_\mathrm{u}$ spanning from $\sim$23\,K to $\sim$1,580\,K and Einstein A coefficient $A_\mathrm{ul}$ of $\sim10^{-8}$--$10^{-6}$), as listed in Table \ref{tab:transitions}.    

The initial calibration of the data was performed using the VLA calibration pipeline within CASA 6.6.1 pipeline 2024.1.1.22\footnote{\url{https://science.nrao.edu/facilities/vla/data-processing/pipeline}}. The two observations were calibrated in separate executions of the calibration pipeline, which performs automated flagging, bandpass calibration, flux density bootstrapping, and standard gain calibration. We then ran the VLA imaging pipeline\footnote{\url{https://science.nrao.edu/facilities/vla/data-processing/pipeline/vipl}} \revise{for} both observations together, which includes continuum finding and subtraction \revise{(with \texttt{hif\_findcont and hif\_uvcontsub} where \texttt{fitorder}\,$=1$)}, automated self-calibration, and imaging of the cubes.



For subsequent imaging of continuum and lines, we used the modular version of the Common Astronomy Software Applications (CASA; \citealt{CASA}) version 6.6.0.20. We first imaged the calibrated visibilities to generate the continuum image by the \texttt{tclean} task with a robust parameter of 0.5, which resulted in a beam size of $0\farcs53\times0\farcs32$ (P.A. $=-4\arcdeg$) and a RMS noise level of 7.7\,\textmu Jy beam$^{-1}$.
We then imaged the \revise{continuum-subtracted} spectral line visibilities for each transition listed in Table \ref{tab:transitions} by the \texttt{tclean} task with a robust parameter of 1.0 in the modified Briggs weighting \texttt{briggsbwtaper} and a channel width of 0.8 km s$^{-1}$. \revise{These choices achieve the best balance between the line sensitivity and spatial/spectral resolutions.} The resulting beam size and rms noise level \revise{(both in mJy beam$^{-1}$ and K)} for each image cube are reported in Table \ref{tab:transitions}\revise{, where the rms noise levels are measured on the emission-free channels}. \revise{We convert the rms values to the brightness temperature scale using the Rayleigh-Jeans approximation, and this applies throughout the paper where the brightness temperature unit is used.} We also generated the beam-matched image cubes for all transitions with a beam size of $0\farcs8\times0\farcs5$ (P.A. $=0\arcdeg$) by smoothing each image cube using the \texttt{imsmooth} task in CASA. 

\begin{deluxetable*}{cCCCRLcc}
\label{tab:transitions}
\tablecaption{Properties of the Observed NH$_3$ Transitions}
\tablehead{\colhead{Transition} & \colhead{$\nu_0$} & \colhead{$\log_{10}A_\mathrm{ul}$} & \colhead{$g_\mathrm{u}$} & \colhead{$E_\mathrm{u}$} & \colhead{Beam (P.A.)} & \multicolumn{2}{c}{RMS} \\ 
\colhead{$(J, K)$} & \colhead{(GHz)} & \colhead{(s$^{-1}$)} & \colhead{ } & \colhead{(K)} & \colhead{ } & \colhead{(mJy beam$^{-1}$)} & \colhead{(K)}}
\startdata
(1, 1) & 23.694496 & -6.7753 & 6 & 23.26 & 0\farcs60\times0\farcs36\,(-2\fdg0) & 1.2 & 12 \\
(2, 2) & 23.722633 & -6.6490 & 10 & 64.45 & 0\farcs60\times0\farcs36\,(-2\fdg1) & 1.1 & 11 \\
(3, 3) & 23.870129 & -6.5901 & 28 & 123.54 & 0\farcs60\times0\farcs36\,(-1\fdg6) & 1.1 & 11 \\
(4, 4) & 24.139416 & -6.5482 & 18 & 200.52 & 0\farcs61\times0\farcs35\,(-0\fdg68) & 1.3 & 13 \\
(5, 5) & 24.532990 & -6.5101 & 22 & 295.37 & 0\farcs58\times0\farcs35\,(-0\fdg94) & 1.1 & 11 \\
(5, 4) & 22.653023 & -6.8059 & 22 & 343.31 & 0\farcs65\times0\farcs37\,(-1\fdg5) & 1.4 & 14 \\
(6, 6) & 25.056025 & -6.4714 & 52 & 408.06 & 0\farcs57\times0\farcs34\,(-0\fdg63) & 1.1 & 11 \\
(5, 1) & 19.838346 & -8.1797 & 22 & 422.77 & 0\farcs71\times0\farcs43\,(-1\fdg5) & 1.0 & 10 \\
(7, 7) & 25.715181 & -6.4297 & 30 & 538.55 & 0\farcs54\times0\farcs32\,(1\fdg0) & 1.2 & 12 \\
(6, 3) & 19.757538 & -7.3773 & 52 & 551.32 & 0\farcs71\times0\farcs43\,(-1\fdg2) & 1.0 & 11 \\
(6, 2) & 18.884695 & -7.7873 & 26 & 577.65 & 0\farcs75\times0\farcs45\,(-0\fdg59) & 1.0 & 10 \\
(8, 6) & 20.719221 & -6.9498 & 68 & 835.47 & 0\farcs68\times0\farcs43\,(3\fdg1) & 1.1 & 10 \\
(9, 7) & 20.735453 & -6.9126 & 38 & 1022.62 & 0\farcs68\times0\farcs43\,(2\fdg8) & 1.1 & 11 \\
(10, 8) & 20.852528 & -6.8775 & 42 & 1227.44 & 0\farcs67\times0\farcs43\,(2\fdg5) & 1.1 & 11 \\
(11, 10) & 24.881922 & -6.5377 & 46 & 1349.57 & 0\farcs57\times0\farcs34\,(-0\fdg95) & 1.0 & 10 \\
(11, 9) & 21.070740 & -6.8420 & 92 & 1449.88 & 0\farcs67\times0\farcs44\,(-0\fdg38) & 1.3 & 12 \\
(12, 11) & 25.695230 & -6.4873 & 50 & 1579.06 & 0\farcs55\times0\farcs33\,(0\fdg20) & 1.1 & 11
\enddata
\tablecomments{\textrm{The spectroscopic data are taken from the Jet Propulsion Laboratory (JPL) database \citep{JPL}. The original data are presented in \citet{Yu2010}.}}
\end{deluxetable*}

\section{Observational Results} \label{sec:results}

\subsection{1.2 cm Continuum}\label{subsec:continuum}
\begin{figure}
\plotone{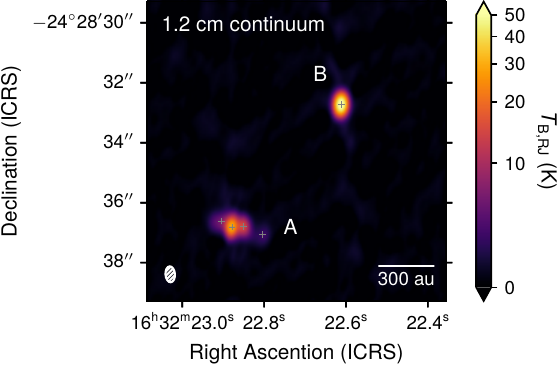}
\caption{The 1.2\,cm continuum image of IRAS 16293-2422. We note that the color scales employ arcsinh stretches, with the lower end saturating at 0.0. \revise{The gray crosses mark the peaks of emission components estimated from the Gaussian fit as described in Section \ref{subsec:continuum}. For source A, the crosses correspond to the component 4 (A2$\beta$), 1 (A1), 2 (A2), and 3 (A2$\alpha$), from east to west (see Table \ref{tab:gaussian_fit}).}}
\label{fig:continuum}
\end{figure}

Figure \ref{fig:continuum} shows the 1.2\,cm continuum image of IRAS 16293-2422. 
Two emission components associated with source A and B are clearly resolved. While source B appears as compact, source A \revise{is} highly structured with four possible emission components, whose overall morphology is consistent with the previous VLA observations \citep[e.g.,][]{Hernandez-Gomez2019}. 

To characterize the properties of the continuum emission, we performed the two-dimensional Gaussian fit to the continuum image using the CASA task \texttt{imfit}. For source B, we considered a single component, while four components were taken into account for source A. The results of the fits, including peak coordinates, deconvolved sizes, peak intensities, and flux densities, are reported in Table \ref{tab:gaussian_fit}. For source A, the two brightest components (component 1 and 2) corresponds to the confirmed protostars A1 and A2 based on the high-resolution kinematics of CS and COM lines \citep{Maureira2020}, while other two components are thought to be ejecta (A2$\alpha$ and A2$\beta$)\revise{, likely associated with the outflow originated from the protostar A2} \citep{Loinard2007, Hernandez-Gomez2019}. The peak coordinates of the brightest components (component 1) for both source A and B are used as the source position in the following analyses.

\begin{deluxetable*}{lcccc}
\tablecaption{Gaussian Fit to the Continuum Image}
\label{tab:gaussian_fit}
\tablehead{\colhead{Component} & \colhead{Peak Coordinate (ICRS)} & \colhead{Deconvolved Size (P.A.)} & \colhead{Peak Intensity} & \colhead{Flux Density} \\
\colhead{} & \colhead{} & \colhead{} & \colhead{(mJy beam$^{-1}$)} & \colhead{(mJy)} 
}
\startdata 
\multicolumn{5}{c}{Source A} \\
\hline
1 (A1) & $16^\mathrm{h} 32^\mathrm{m} 22\fs 879$ $-24^\mathrm{d}28^\mathrm{m}36\fs81$ & $0\farcs18\times0\farcs05$ ($176\arcdeg$) & $46.7\pm0.3$ & $2.53\pm0.03$ \\
2 (A2) & $16^\mathrm{h} 32^\mathrm{m} 22\fs 850$ $-24^\mathrm{d}28^\mathrm{m}36\fs79$ & $0\farcs29\times0\farcs15$ ($49\arcdeg$) & $6.22\pm0.09$ & $1.63\pm0.04$ \\
3 (A2$\alpha$) & $16^\mathrm{h} 32^\mathrm{m} 22\fs 804$ $-24^\mathrm{d}28^\mathrm{m}37\fs07$ & $0\farcs47\times0\farcs21$ ($125\arcdeg$) & $0.77\pm0.06$ & $0.45\pm0.05$ \\
4 (A2$\beta$) & $16^\mathrm{h} 32^\mathrm{m} 22\fs 903$ $-24^\mathrm{d}28^\mathrm{m}36\fs61$ & $0\farcs55\times0\farcs24$ ($96\arcdeg$) & $1.39\pm0.05$ & $1.10\pm0.06$ \\
\hline
\multicolumn{5}{c}{Source B} \\
\hline
1 & $16^\mathrm{h} 32^\mathrm{m} 22\fs 612$ $-24^\mathrm{d}28^\mathrm{m}32\fs73$ & $0\farcs20\times0\farcs16$ ($14\arcdeg$) & $33.1\pm0.1$ & $6.08\pm0.04$ \\
\enddata
\tablecomments{For source A, correspondences to the name of components in the literature (e.g., \citealt{Hernandez-Gomez2019}) are denoted. The uncertainties on the peak intensity and flux density do not include the absolute flux calibration uncertainty ($\sim15$\%).}
\end{deluxetable*}

\subsection{Line Detections}\label{subsec:line_detections}
\begin{figure*}
\plotone{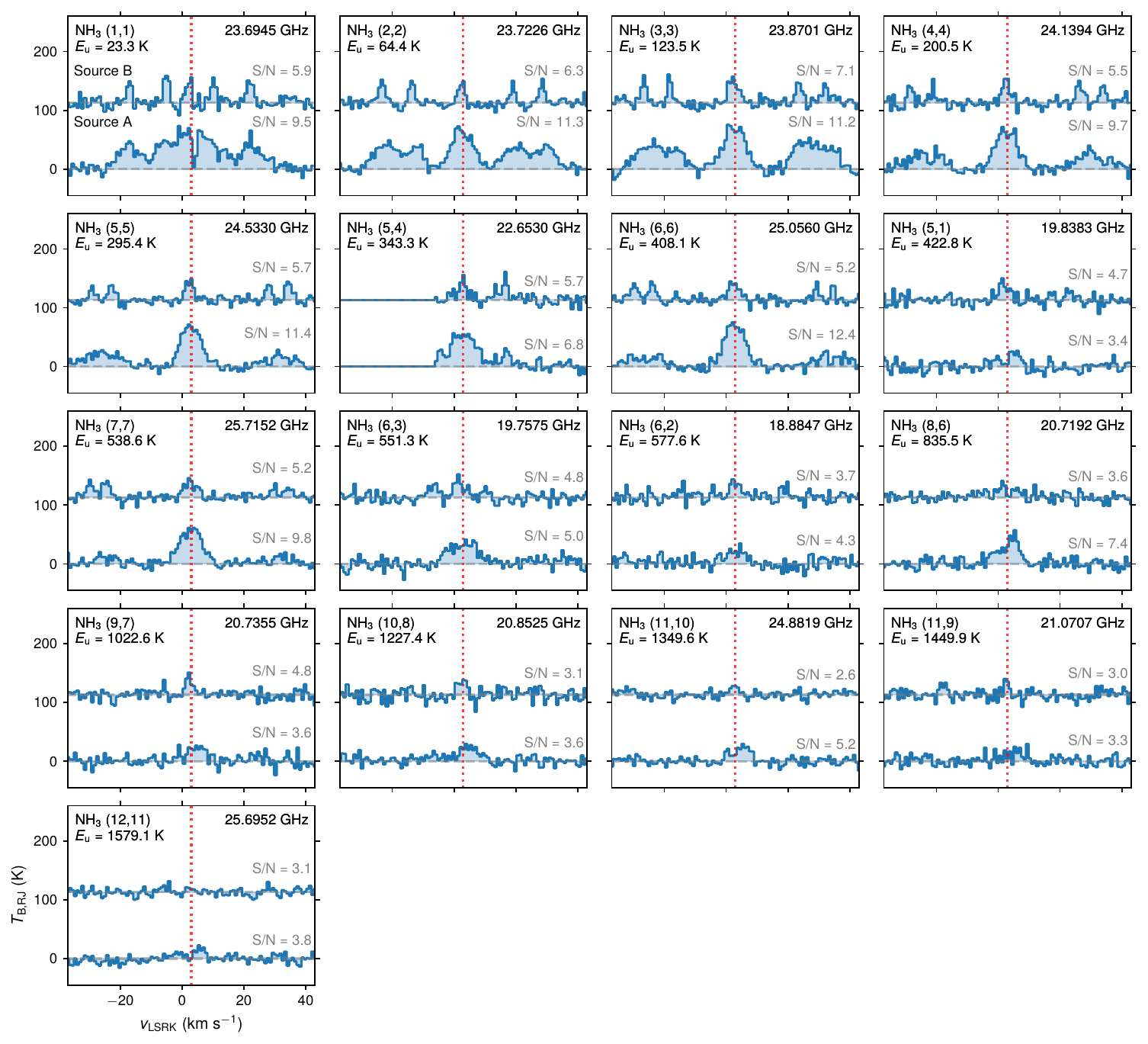}
\caption{Spectra of the observed NH$_3$ transitions at the source positions of source A (bottom) and source B (top). The vertical red dotted line marks the approximate systemic velocity of the source (3 km s$^{-1}$). \revise{Note that the blue-shifted side of the NH$_3$ (5,4) lines are out of the spectral window coverage.}}
\label{fig:spectrum}
\end{figure*}

Figure \ref{fig:spectrum} show the spectra of the observed transitions extracted from the source positions on the beam-matched image cubes. \revise{The peak signal-to-noise ratio (S/N) of each transition is indicated in each panel.} Clear \revise{(S/N $>5$)} spectral signatures of several metastable ($J=K$) and non-metastable ($J\neq K$) NH$_3$ transitions appear in both source A and B, including the prominent hyperfine structure lines particularly for low-$J$ transitions\footnote{\revise{We note that the blue-shifted side of the NH$_3$ (5,4) line is not covered by the spectral window. This does not significantly affect the following analysis as the continuum subtraction is performed accurately enough using the redshifted side, although the noise levels at the edge channels, which are measured on the emission-free region in each channel, are slightly higher than \rrevise{the other lines}.}}. We also see a few high-$J$ lines with high upper state energies ($E_\mathrm{u} > 1000$ K) 
at low S/Ns ($\lesssim3$) for both source A and B. 
To examine the detection significance of these weak lines, we perform \revise{a} line stacking analysis. 
We employed the standard S/N-weighted spectral stacking that should maximize the resulting S/N of the stacked spectra under the assumption of a white noise. We used a subset of NH$_3$ lines with $E_\mathrm{u} > 1000$\,K, namely, ($J, K$) $=$ (9,7), (10,8), (11,9), (11,10), and (12,11) inversion transitions, which are likely share a common spatial origin (see Section \ref{subsec:distributions}).
We treat each transition as a single spectral line ignoring any hyperfine splittings. No hyperfine spectroscopic data are available for the NH$_3$ transitions used here, but they are expected to be weak for high-$J$ lines unless the emission is extremely optically thick.

\begin{figure}
\plotone{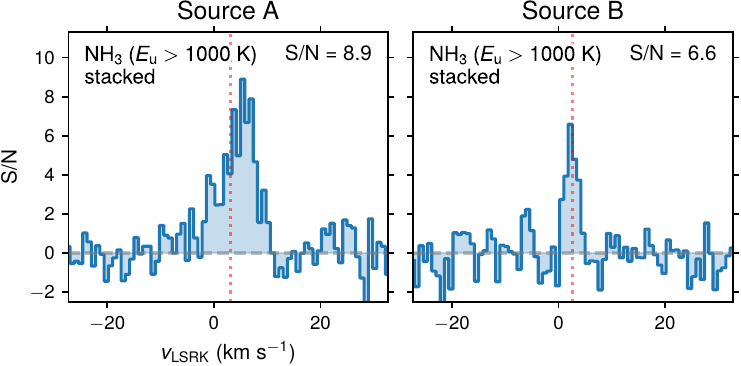}
\caption{Stacked spectra of NH$_3$ transitions with $E_\mathrm{u} > 1,000$\,K in source A (left) and source B (right) that demonstrate the robust detection of high-excitation lines. The intensity scale is normalized by the noise levels of each spectrum. The vertical red dotted line marks the systemic velocity of each source.}
\label{fig:stacked_spectrum}
\end{figure}

Figure \ref{fig:stacked_spectrum} shows the resulting stacked spectra. The S/Ns of the stacked lines reach $\gtrsim7$ in both source A and source B, confirming the detection of these lines. This is the first detection of such high-$E_\mathrm{u}$ ($> 1000$\,K) NH$_3$ lines in low-mass star-forming regions, while they have historically been detected in \revise{high-mass star-forming regions, such as Orion~KL} \citep{hermsen1988, wilson1993, wilson2006} and Sgr~B2 \citep{huttemeister1993}. 

\subsection{Spatial Distributions and Kinematics}\label{subsec:distributions}
\begin{figure*}
\plotone{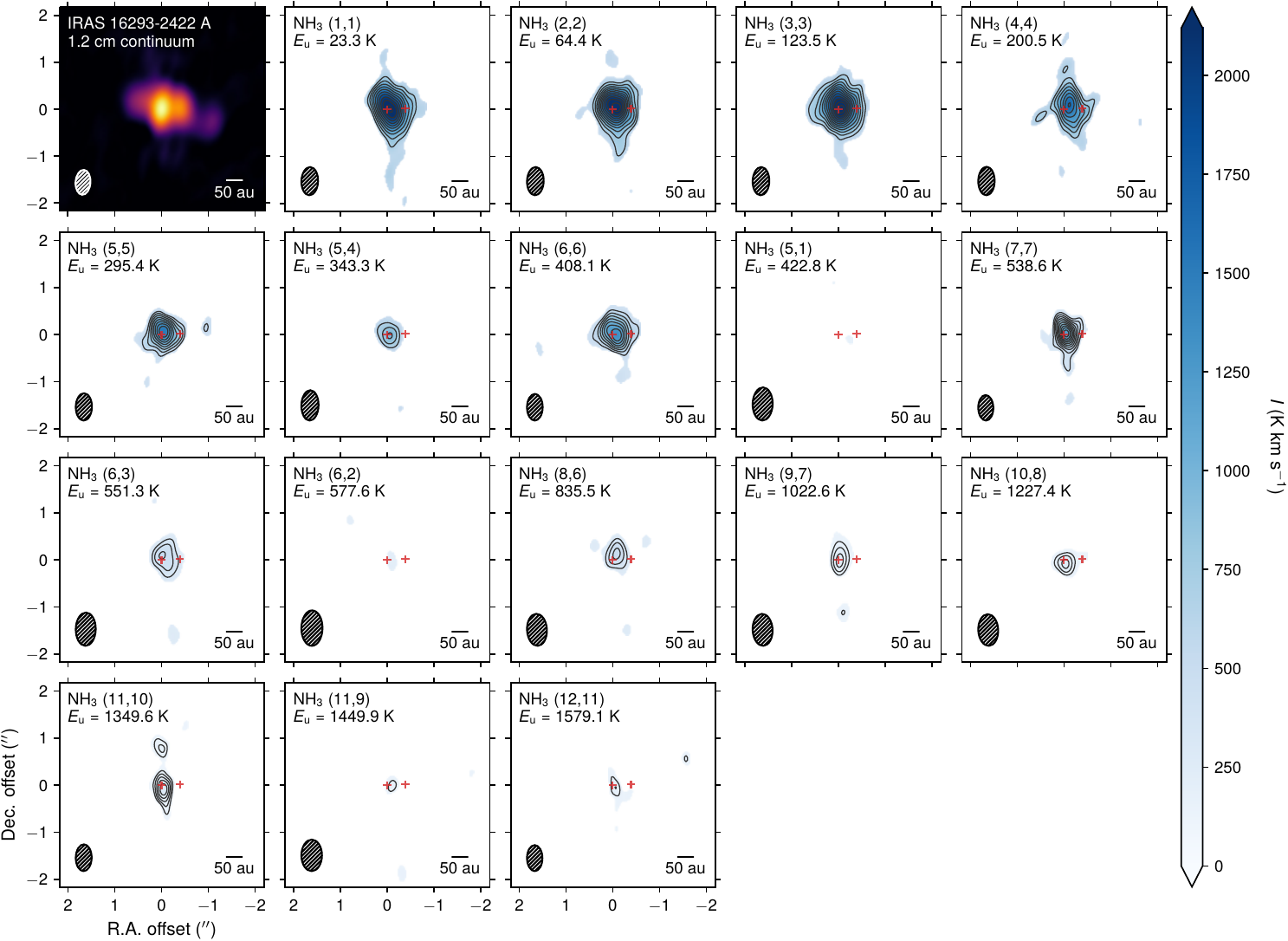}
\caption{1.2\,cm continuum image (top left) and velocity-integrated intensity maps of observed NH$_3$ transitions (others) toward source A. Only pixels with values greater than $3\sigma$ are shown, where $\sigma$ is the rms value measured on the emission-free region of each map. The contours on each panel start from 3$\sigma$ followed by a 1$\sigma$ step. The red crosses indicate the positions of the protostars A1 and A2.}
\label{fig:sourceA_mom0}
\end{figure*}

\begin{figure*}
\plotone{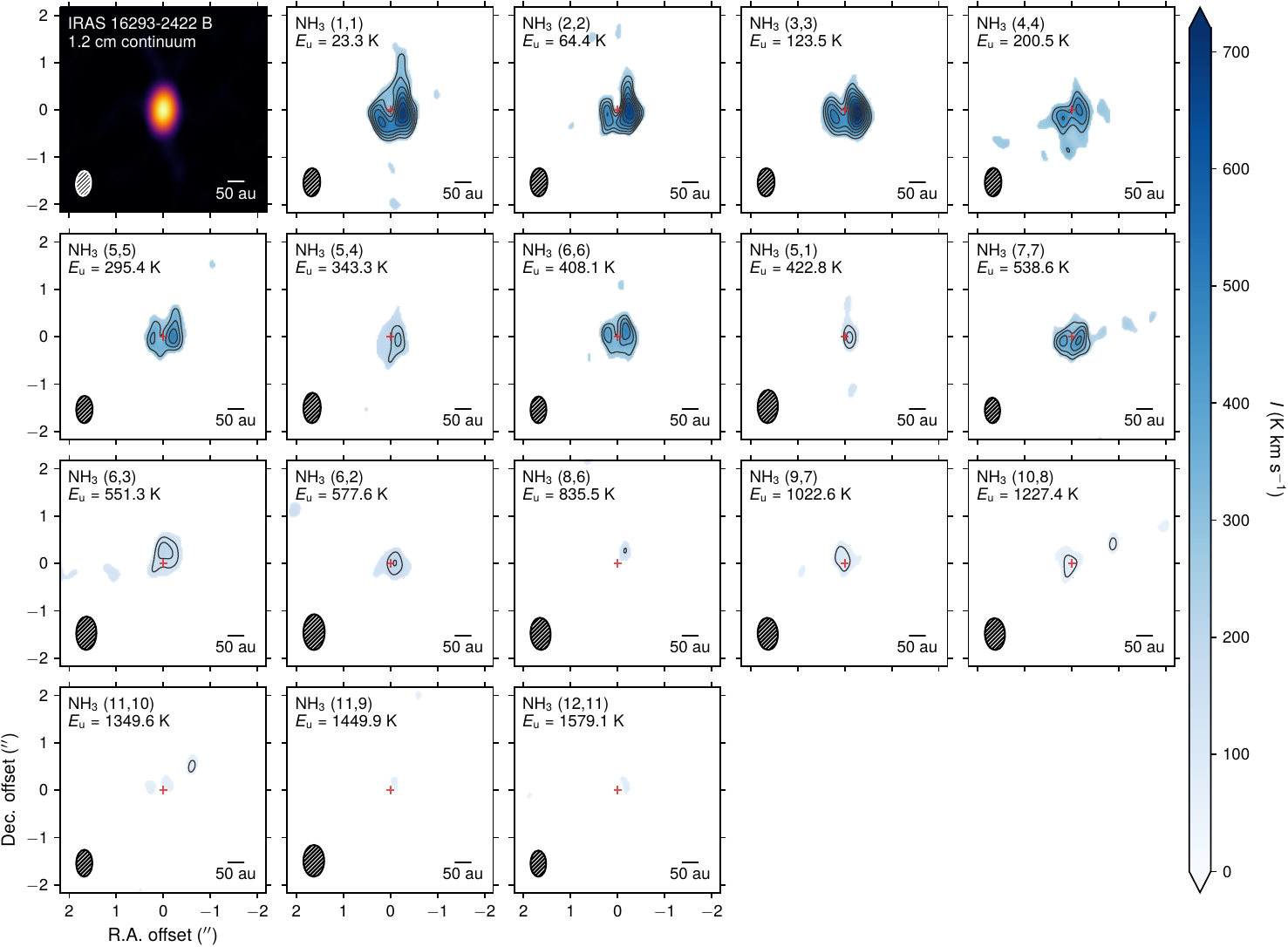}
\caption{Same as Figure \ref{fig:sourceA_mom0}, but for source B.}
\label{fig:sourceB_mom0}
\end{figure*}

To investigate the spatial distributions of the line emission, we generated the velocity-integrated intensity maps of each transition using the Python package \texttt{bettermoments} \citep{bettermoments} with no intensity threshold for pixel inclusion. The velocity ranges for integration are $\pm5$ and $\pm1.5$ km s$^{-1}$ with respect to the systemic velocity (3.1 and 2.7 km s$^{-1}$; \citealt{Jorgensen2016}) for source A and B, respectively, where hyperfine splitting are taken into account if the spectroscopic data are available. \revise{These velocity range choices are based primarily on the visual inspection of the spectra, which are further confirmed to be appropriate from the line widths of COMs reported in previous studies \citep[e.g.,][]{Jorgensen2016, Manigand2020}}. Figure \ref{fig:sourceA_mom0} and \ref{fig:sourceB_mom0} show the maps of each observed transition for source A and source B respectively. Most of the NH$_3$ line emission is spatially resolved. For source A, the NH$_3$ emission overall appears as originating from the circumbinary region with peaks at the A1 protostar position. The high-$J$ lines with $E_\mathrm{u}>1000$\,K show more compact distributions at A1 compared to the low-$J$ lines, indicating that they selectively trace the inner hot region. 
For source B, there are notable differences in spatial distributions between the low-$J$ lines, namely NH$_3$ (1,1), (2,2), and (3,3), and the higher-$J$ lines; low-$J$ lines show the peaks offset from the center to the west, while higher-$J$ lines are more centrally concentrated. 

To highlight these varying spatial distributions depending on the line excitation properties in both source A and source B, we present the stacked velocity-integrated intensity maps and their spatial profiles for each low-$J$ ($E_\mathrm{u} < 150$\,K) and high-$J$ ($E_\mathrm{u} > 1000$\,K) lines in Figure \ref{fig:comparison_stacked_highE_lowEu}, whereas the stacked channel maps for each can also be found in Appendix \ref{appendix:channel_maps}. The stacked maps were generated by stacking each set of maps weighted by peak S/Ns. Prior to stacking, the maps were smoothed to the smallest achievable common beam (roughly $\sim0\farcs8\times0\farcs5$). The spatial profiles were computed along P.A. $=90\fdg0$, which is almost the direction of beam minor axis and for source A is passing through two protostars A1 and A2 with A1 centered. It is clear for both source A and source B that the high-$J$ lines show a compact emission compared to the low-$J$ lines. The difference in spatial distribution is more pronounced in source B, in which the high-$J$ lines are peaking at the center (position of the protostar) while low-$J$ lines show the depression. The distribution of the low-$J$ lines in source B shows a emission peak at the west side of the protostar with $\sim0\farcs3$ offset, which is consistent with the distributions of the COM emission observed with ALMA at a comparable resolution \citep[e.g.,][]{Oya2018}.

The top panels of Figure \ref{fig:peak_Tb_velocity} shows the peak brightness temperature map of NH$_3$ (3,3) in both sources. The line brightness temperature is calculated using \texttt{bettermoments} package \citep{bettermoments}, to which the continuum brightness temperature is subsequently added to form the line plus continuum brightness temperature maps. 
\revise{The observed hyperfine intensity ratios (satellite to main) of NH$_3$ (3,3) reach $\sim$0.5 for source A and nearly unity for source B (Figure \ref{fig:spectrum}), both of which are higher than the intrinsic intensity ratio of $0.03$ (see Table \ref{tab:transitions_hfs}), indicating that the line is highly optically thick. Thus},
the line plus continuum brightness temperature is a proxy of the gas temperature \citep{Weaver2018, Law2021}. 
The brightness temperatures are $\gtrsim100$\,K at the inner $\sim0\farcs5$ region for both source A and source B, but with a slightly lower temperature in source B, likely due to its smaller source size and thereby severe beam dilution effect. The high temperature ($>100$\,K) in the vicinity of protostars is broadly consistent with the \revise{previous estimates using multi-line analysis of} COMs \citep{Oya2016, Jorgensen2016, Oya2018} and H$_2$CS \citep{vantHoff2020, Oya2020}.

The intensity-weighted velocity (moment 1) maps of NH$_3$ (3,3), which are produced using \texttt{bettermoments} package, are shown in the bottom panels of Figure \ref{fig:peak_Tb_velocity}. \revise{The maps are produced using a velocity range of $\pm10$\,km\,s$^{-1}$ with respect to the systemic velocities (encompassing only the main hyperfine component) and an intensity threshold of $\geq3.5\sigma$ for pixel inclusion.} While the emission in source A clearly shows velocity gradient along the northeast--southwest direction, no strong velocity gradients are observed in source B \revise{at the current spatial/spectral resolution}. These characteristics are consistent with the ALMA observations of CH$_3$OH and HCOOCH$_3$ lines that trace the similar spatial scales \citep{Oya2016, Oya2018, Maureira2020}. On the other hand, the high-$E_\mathrm{u}$ ($> 1,000$\,K) lines show different velocity structures, particularly in source A; these lines show no clear velocity gradients at the current resolution but stronger redshifted emission as indicated in the stacked spectra shown in the left panel of Figure \ref{fig:stacked_spectrum}. 

In addition, we found a hint of an asymmetric line profile for a subset of observed lines in source B.
Figure \ref{fig:stacked_spectrum_infall} shows the stacked line profile of NH$_3$ (5,1), (7,7), (6,3), (6,2), (8,6), and (9,7). \revise{These transitions are selected based on the upper state energies ($\approx$400--1000\,K), where the spectra of these transitions exhibit slight shifts of the peak velocity (Figure \ref{fig:spectrum}).}
Although the S/N even after stacking is not highly convincing, the peak velocity is slightly blueshifted, forming a so-called blue-skewed profile. We will discuss the interpretation of this in Section \ref{subsubsec:sourceB}.

\begin{figure*}
\plotone{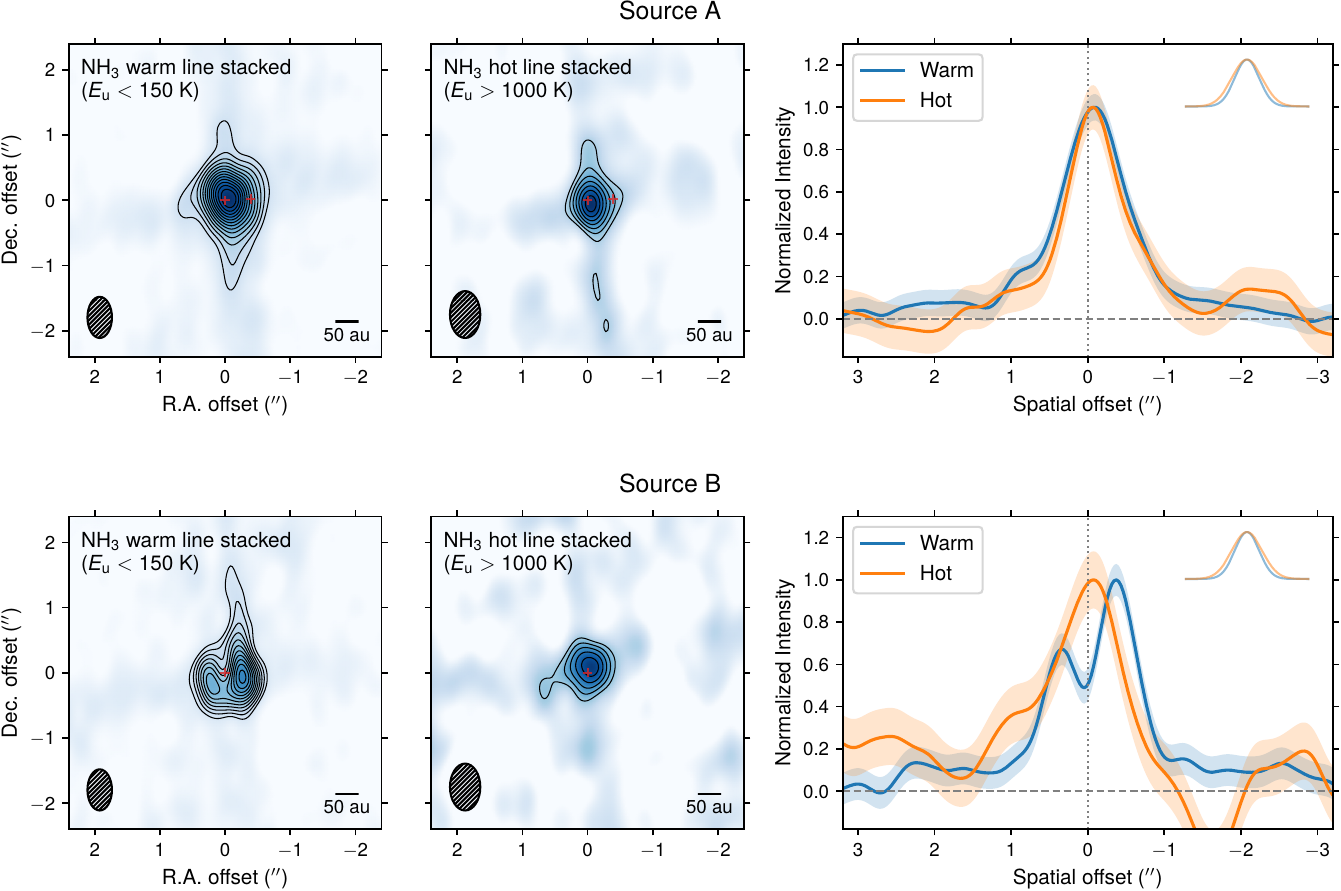}
\caption{Stacked velocity-integrated intensity maps (left) for each set of ``warm'' lines ($E_\mathrm{u}<150$\,K) and ``hot'' lines ($E_\mathrm{u}>1000$\,K), and their normalized spatial profiles along P.A. $=90\fdg0$ (along the Right Ascension or beam minor axis; right). Beam size of the stacked maps and the scale bar of 50\,au are indicated at the lower left and lower right of each map, respectively. The contours start from 3$\sigma$ followed by $1\sigma$ step. The red crosses indicate the position of protostars. The color-shaded regions of the spatial profiles indicate $1\sigma$ uncertainty. The Gaussian profile at the upper right of the right panels \revise{are the} beam profile along its minor axis.}
\label{fig:comparison_stacked_highE_lowEu}
\end{figure*}

\begin{figure}
\epsscale{1.1}
\plotone{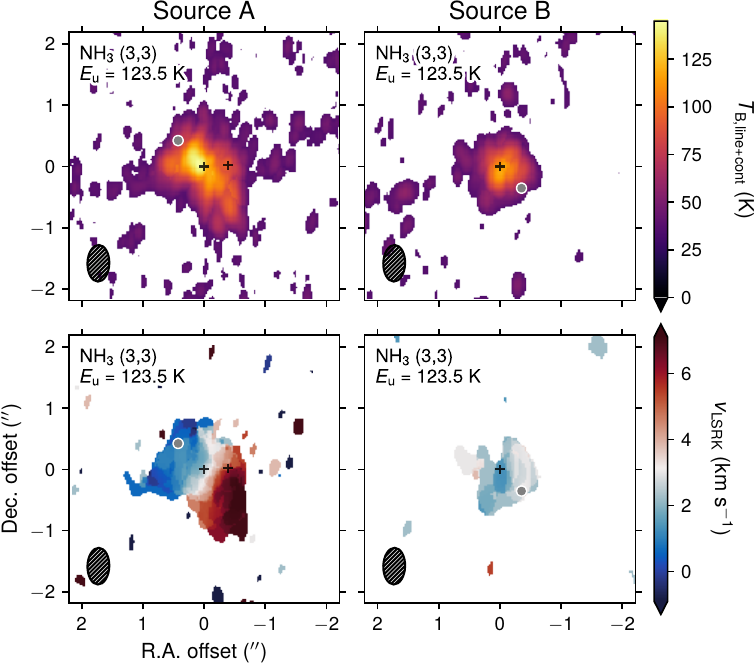}
\caption{Line plus continuum peak brightness temperature (top) and intensity-weighted velocity (bottom) maps of the NH$_3$ (3,3) emission in source A (left) and source B (right). The brightness temperature is converted assuming the Rayleigh-Jeans approximation. Only pixels with a peak S/N $\geq3$ are shown for the brightness temperature maps. The black crosses indicate the positions of protostars\revise{, while the gray circles mark the offset positions used for the spectral fits described in Section \ref{subsec:fit_offset}}.}
\label{fig:peak_Tb_velocity}
\end{figure}

\begin{figure}
\epsscale{0.8}
\plotone{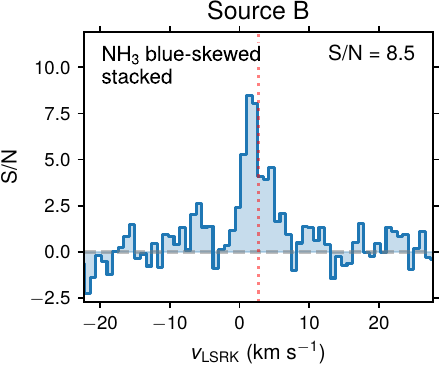}
\caption{Stacked spectrum of NH$_3$ (5,1), (7,7) (6,3), (6,2), (8,6), and (9,7) lines toward source B, which shows a blue-skewed line profile.}
\label{fig:stacked_spectrum_infall}
\end{figure}

\section{Analysis}\label{sec:analysis}
To characterize the properties of the observed NH$_3$ gas around the protostars, we employed excitation analyses 
taking advantage of the multi-line nature of the data.
In Section \ref{subsec:fit_offset}, we employed spectral fits at the offset positions from the protostars: 0\farcs6 north-east and 0\farcs5 south-west offset positions for source A and source B, respectively\revise{, as indicated in Figure \ref{fig:peak_Tb_velocity}}. These are the same positions as used in the analysis of the PILS data to avoid the severe line blending and absorption \citep[e.g.,][]{Jorgensen2018, Manigand2020}. In Section \ref{subsec:fit_protostar}, we fit the velocity-integrated intensities at the protostar positions of source A and source B to constrain the properties of the high-$J$ line emission that is clearly detected only in the vicinity of protostars. 
Throughout the analysis described below, we assume that the excitation is thermalized, namely, in the local thermodynamical equilibrium (LTE) condition, given the high density \revise{($\sim$$10^7$--$10^8$\,cm$^{-3}$; e.g., \citealt{Crimier2010, Jacobsen2018})} of the inner envelope. This assumption is found to be reasonable for the CH$_3$OH lines \citep{Jorgensen2016}, which have similar critical densities \revise{($\sim$$10^3$--$10^4$\,cm$^{-3}$; e.g., \citealt{Shirley2015})} and show a similar emission distribution to the NH$_3$ lines presented here. We also assume that the ortho-to-para ratio of NH$_3$ is unity, which is the
statistical value expected under the warm ($\gtrsim100$\,K) condition. This assumption is also used and found to be valid in the previous NH$_3$ observations of hot corinos in the NGC1333 IRAS4 region \citep{Yamato2022, DeSimone2022}.

\subsection{Offset Positions}\label{subsec:fit_offset}
To fit the spectra extracted from the offset positions, we used a simple slab model similar to that used in the analysis of the PILS data as described in e.g., \citet{Manigand2020}. We assume that the observed NH$_3$ emission can be described by a single emitting component. This assumption is justified from the similar spatial distributions among different low-$J$ ($E_\mathrm{u} < 1,000$\,K) transitions (Figures \ref{fig:sourceA_mom0} and \ref{fig:sourceB_mom0}). Taking the hyperfine splitting into account, the optical depth profile of each inversion transition is
\begin{equation}\label{eq:optical_depth}
    \tau (v) = \sum_{i}r_i\tau_0\exp\left[-\frac{(v - v_0 - \delta v_i)^2}{2\sigma_v^2}\right],
\end{equation}
where the summation $i$ goes over all the hyperfine components, $r_i$ and $\delta v_i$ are the relative intensity and velocity offset of each hyperfine component as listed in Table \ref{tab:transitions_hfs}, $\tau_0$ is the total optical depth of the inversion transition, $v_0$ is the line center velocity, and $\sigma_v$ is the velocity line width. Note that the hyperfine spectroscopic data are available only for subset of inversion transitions (Table \ref{tab:transitions_hfs}), and we treat the transitions without these data as single lines. The total optical depth $\tau_0$ and the emergent intensity for each inversion transition can be expressed as
\begin{equation}\label{eq:tau0}
    \tau_0 = \frac{c^2g_\mathrm{u}A_\mathrm{ul}N}{8\pi\nu_0^2\sqrt{2\pi}\sigma_\nu Q(T)}\left[\exp\left(\frac{h\nu_0}{k_\mathrm{B}T}\right) - 1\right]\exp\left(-\frac{E_\mathrm{u}}{k_\mathrm{B}T}\right),
\end{equation}
\begin{equation}\label{eq:sythetic_spectra}
    I_\nu(v) = \eta B_\nu(T)(1 - e^{-\tau(v)}),
\end{equation}
where $\nu_0$, $g_\mathrm{u}$, $A_\mathrm{ul}$, and $E_\mathrm{u}$ are the rest frequency, statistical weight of the upper state, Einstein A coefficient for the spontaneous emission, and the upper state energy as listed in Table \ref{tab:transitions}, $T$ is the excitation temperature (same as the gas temperature here under the assumption of LTE), $N$ is the NH$_3$ column density, $Q(T)$ is the partition function, and $\eta$ is the beam dilution factor. The partition function is taken from the JPL database. The synthetic spectra $I_\nu$ are first generated with a finer channel width that samples $\approx5$ elements across the line width, and then convolved with the actual channel width (0.8 km\,s$^{-1}$) and resampled at the observed velocity sampling. We note that no corrections for the continuum emission are included, given that\revise{, in contrast to the PILS analysis, }the continuum emission are negligibly weak at the offset positions for both source A and source B in the present observations \revise{due to the difference in the observing wavelengths \citep{Hernandez-Gomez2019}.}

\begin{figure*}
\plotone{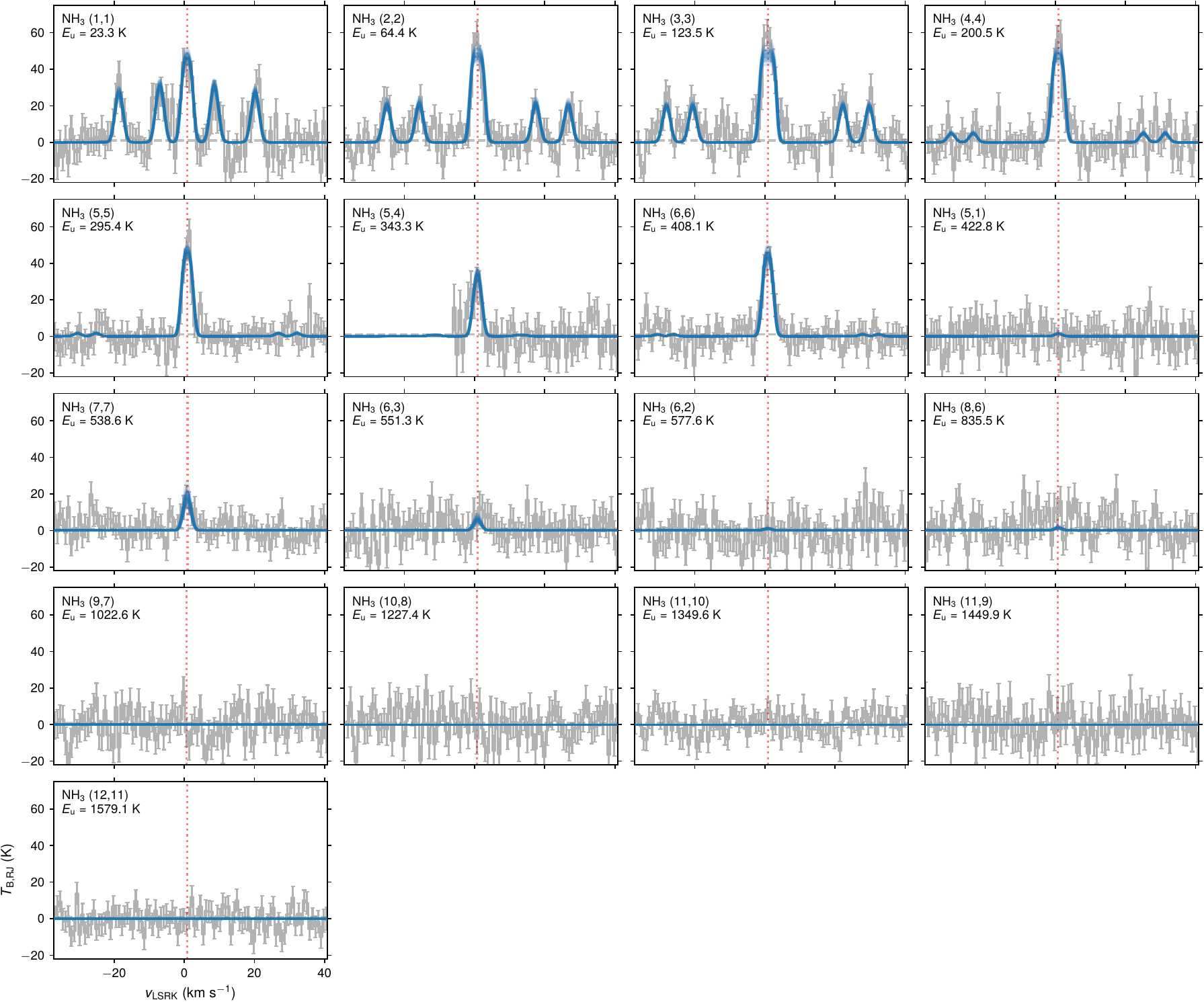}
\caption{Observed (gray) and modeled (blue) spectra at the offset positions of source A (0\farcs6 north-east). The modeled spectra are drawn for 100 samples randomly selected from the posterior chains. The vertical red dotted line marks the best-fit (median of the posterior sample) line center velocity.}
\label{fig:fit_offset_A}
\end{figure*}

\begin{figure*}
\plotone{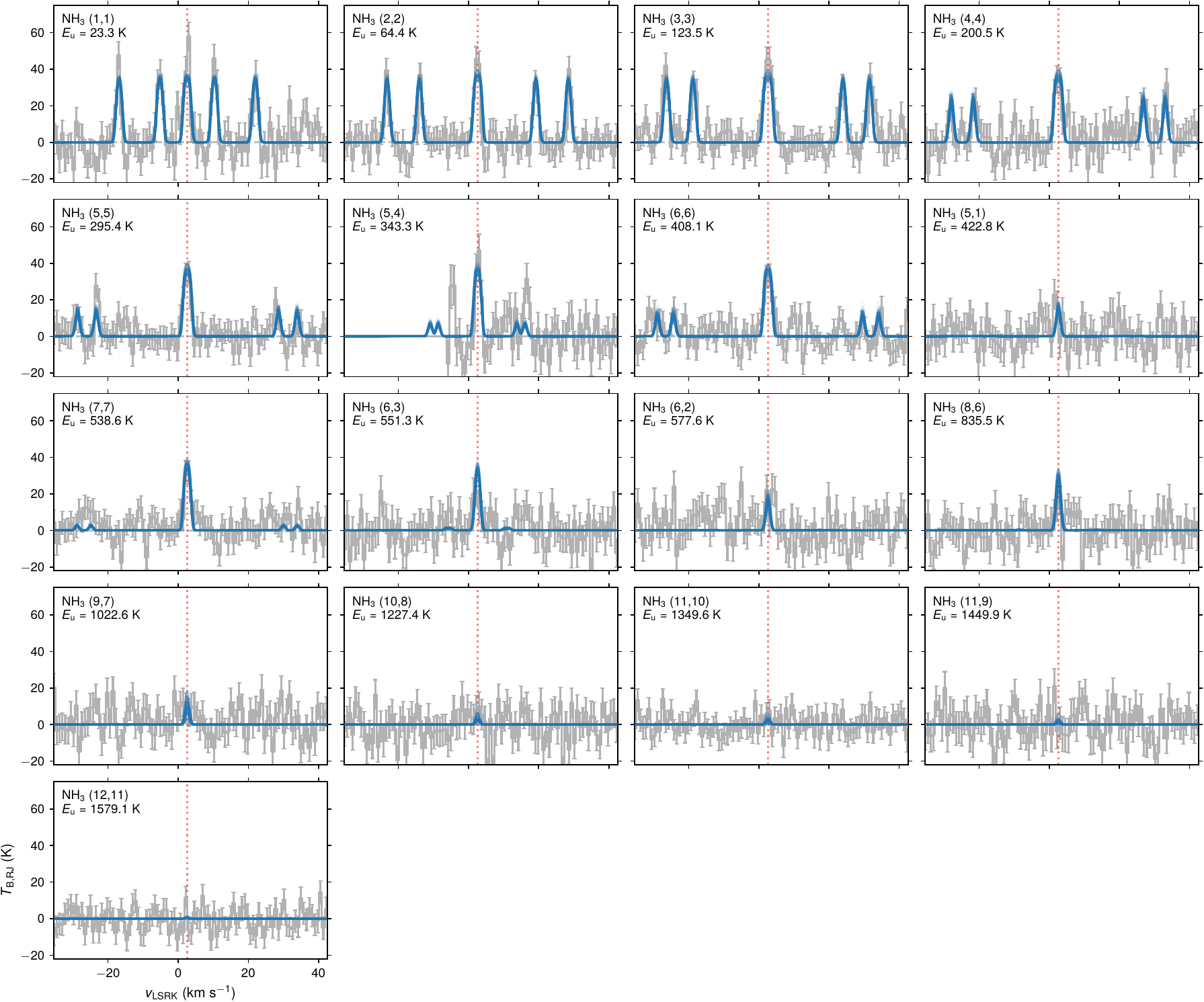}
\caption{Same as Figure \ref{fig:fit_offset_A}, but for source B (0\farcs5 south-west offset position). \revise{Note that a tentative ($\sim3$--$4\sigma$) feature at $\sim17$\,km\,s$^{-1}$ of the NH$_3$ (5,4) transition could either be just a noise spike or a signal of hyperfine satellite that is imperfectly accounted for by the model.}}
\label{fig:fit_offset_B}
\end{figure*}

We fit the synthetic spectra to the observed spectra of all transitions simultaneously, leaving $\eta$, $N$, $T$, $\Delta V_\mathrm{FWHM}$, and $v_0$ as free parameters, where $\Delta V_\mathrm{FWHM} = \sqrt{8\ln 2}\sigma_v$ is the velocity width in full width at half maximum (FWHM). We included a 15\% systematic flux calibration uncertainty \revise{within the likelihood function}.
We used the Markov Chain Monte Carlo (MCMC) algorithm implemented in the Python package \texttt{emcee} \citep{emcee} to explore the parameter space, employing 50 walkers with 2,000 steps with the initial 1,000 steps discarded as burn-in. Figure \ref{fig:fit_offset_A} and \ref{fig:fit_offset_B} show the comparison of the observed and modeled spectra for source A and source B, respectively. At the offset positions, almost no emission of the high-$J$ ($E_\mathrm{u} > 1,000$\,K) lines are present, and all the spectra are well reproduced with this single-component slab model. We note that while there remains optically thin lines that effectively constrain the column density, some lines are highly optically thick ($\tau_0\gg1$), especially the low-$J$ lines in source B, as indicated by the saturated hyperfine satellites. This may result in the underestimation of column density and temperature if, for example, the effect of self-absorption due to the extended cold component is significant and the line intensity is thus reduced. However, for source A, this effect should be negligible as the velocity is significantly offset from the source systemic velocity where the extended component mainly emits. For source B, we consider that the effect of cold extended component is minimal, given that the NH$_3$ (1,1) and (2,2) lines, which could be affected by such effect, also show similar spatial distributions and almost the same peak brightness temperature as that of the optically thick NH$_3$ (3,3) line ($E_\mathrm{u}= 124$\,K) that should not be affected by the cold component. 
The resulting constraints on the parameters are reported in Table \ref{tab:fit_offset}\revise{, where the uncertainties are quoted as 16th--84th percentiles of the posterior MCMC chains}. For both sources, the high excitation temperatures of $\gtrsim100$\,K indicate that the observed emission traces the warm gas around the protostars. The line widths and line center velocities are consistent with the PILS studies \citep[e.g.,][]{Jorgensen2018, Manigand2020}. There are notable differences between source A and source B in the beam dilution factor and column density, where source B has a smaller filling factor and higher column density. This is consistent with more compact distributions and stronger hyperfine satellites in source B given that in general the stronger hyperfine satellites indicate higher optical depth and thereby larger column density.

\begin{deluxetable}{lccccc}
\tablecaption{Constraints on the parameters from the spectral fits toward the offset positions}
\label{tab:fit_offset}
\tablehead{\colhead{Source} & \colhead{$\eta$} & \colhead{$T$} & \colhead{$N$} & \colhead{$\Delta V_\mathrm{FWHM}$} & \colhead{$v_0$} \\
\colhead{} & \colhead{} & \colhead{(K)} & \colhead{(cm$^{-2}$)} & \colhead{(km s$^{-1}$)} & \colhead{(km s$^{-1}$)}
}
\startdata 
A & $0.46_{-0.04}^{+0.04}$ & $104_{-5}^{+5}$ & $3.3_{-0.4}^{+0.5}\times10^{17}$ & $2.3_{-0.1}^{+0.1}$ & $0.81_{-0.06}^{+0.06}$ \\
B & $0.25_{-0.03}^{+0.03}$ & $146_{-15}^{+15}$ & $2.5_{-0.4}^{+0.5}\times10^{18}$ & $1.1_{-0.1}^{+0.1}$ & $2.57_{-0.04}^{+0.04}$ \\
\enddata
\end{deluxetable}

\subsection{Protostar Positions}\label{subsec:fit_protostar}
At the protostar positions, both source A and source B exhibit complex velocity structure associated with the rotating and/or infalling motion of the gas. Given that these components are not fully spatially resolved and the limited S/Ns of lines, we employ the modeling of velocity-integrated intensities rather than line profiles to constrain the properties of the NH$_3$ gas. We compute the velocity-integrated intensities at the protostar positions by spectrally integrating over $\pm 6$ km s$^{-1}$ and $\pm 1.5$ km s$^{-1}$ with respect to the line center for source A and source B, respectively, treating hyperfine components separately if spectrally resolved (i.e., no overlaps of integration ranges). This allows us to accurately and independently estimate the optical depth of each inversion transition from the hyperfine line ratios. We consider two emitting components of the NH$_3$ gas in front of the dust continuum, namely, the warm gas in the circumstellar region and the hot gas at the vicinity of the protostars as traced by the low-$J$ lines and high-$J$ lines, respectively. For each component, we assume a single isothermal slab as formulated in Equation (\ref{eq:sythetic_spectra}). The total modeled intensity is given by
\begin{equation}
    I \approx \left[I_{\nu,\mathrm{warm}} + (I_{\nu, \mathrm{hot}} + I_{\nu, \mathrm{cont}}e^{-\tau_\mathrm{hot,0}})e^{-\tau_\mathrm{warm,0}} - I_{\nu, \mathrm{cont}}\right] \Delta V, \label{eq:two-component_model}
\end{equation}
where $I_{\nu,\mathrm{warm}}$ and $I_{\nu, \mathrm{hot}}$ are the intensity from the warm and hot components respectively, $I_{\nu,\mathrm{cont}}$ is the observed continuum intensity (15\,K and 27\,K in Rayleigh-Jeans brightness temperature for source A and source B, respectively), $\Delta V$ is the line width, and $\tau_\mathrm{warm,0}$ and $\tau_\mathrm{hot,0}$ are the line-center optical depth for warm and hot components, respectively. The final term corresponds to the continuum subtraction. We assume a line width of $\Delta V = 6$ km s$^{-1}$ and 3 km s$^{-1}$ for both components in source A and source B, respectively, based on the visual inspection of the spectra. The number of parameters are in total six for each source; column densities $N$, rotation temperature $T_\mathrm{rot}$, and the beam dilution factor $\eta$ for both warm and hot components. For source A, the beam dilution factor $\eta$ of the warm component is fixed to unity given that the emission of the low-$J$ transitions is mostly spatially resolved (see Figure \ref{fig:sourceA_mom0}). 

\begin{figure*}
\plotone{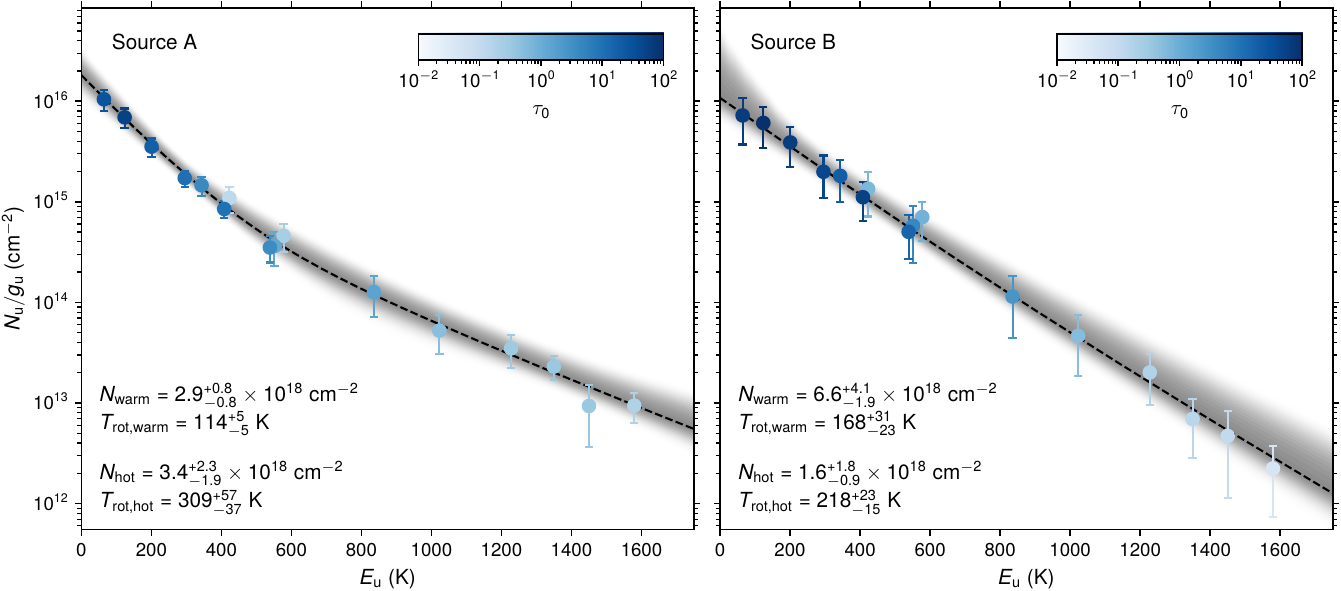}
\caption{Result of the model fit to the velocity-integrated intensities at the protostar positions of source A (left) and source B (right) in the form of rotation diagram. Random drawn from the posterior are shown in gray and the best-fit model that maximize the likelihood is plotted by the black dashed curve. The color represents the maximum optical depth among the two components for each inversion transition.}
\label{fig:fit_peak}
\end{figure*}

We fit this model to the velocity-integrated intensities of all the observed transitions except NH$_3$ (1,1) simultaneously to constrain these parameters. We exclude NH$_3$ (1,1) from the fit because the NH$_3$ (1,1) emission is likely affected by the absorption due to the cold extended component as indicated by the absorption feature particularly in source A (see Figure \ref{fig:spectrum}). We included a 15\% systematic flux calibration uncertainty during the fit. The parameter space is explored by the MCMC algorithm using \texttt{emcee}, employing 100 walkers with 10,000 steps with the initial 5,000 steps discarded as burn-in. Figure \ref{fig:fit_peak} shows the result in the form of rotation diagram. All the observed line emission is well reproduced by this two-component model. The resulting constraints on the parameters are listed in Table \ref{tab:fit_peak}. For both source A and source B, the column density and rotation temperature are higher than (or comparable to) the values at the offset position (Table \ref{tab:fit_offset}), indicating the denser and warmer condition at the central region of the circumstellar structure. The rotation temperatures of the hot component are even higher for both sources, while the temperature difference between the warm and hot components are smaller in source B. Given this small temperature difference, we also performed a single component fit (i.e, $I_{\nu, \mathrm{hot}} = 0$ and $\tau_\mathrm{hot,0} = 0$ in Equation \ref{eq:two-component_model}) for source B, where we derive an intermediate value of rotation temperature ($\sim200$\,K) and a column density consistent with the sum of those of the two components (Appendix \ref{appendix:peak_single_fit_sourceB}).

\begin{deluxetable}{llccc}
\tablecaption{Constraints on the parameters from the fits toward the protostar positions}
\label{tab:fit_peak}
\tablehead{\colhead{Source} & \colhead{Component} & \colhead{$\eta$} & \colhead{$T_\mathrm{rot}$} & \colhead{$N$} \\
\colhead{} & \colhead{} & \colhead{} & \colhead{(K)} & \colhead{(cm$^{-2}$)} 
}
\startdata
A & Warm & 1.0 (fixed) & $114_{-5}^{+5}$ & $2.9_{-0.8}^{+0.8}\times10^{18}$ \\
  & Hot & $0.34_{-0.08}^{+0.15}$ & $310_{-40}^{+60}$ & $3.4_{-1.9}^{+2.3}\times10^{18}$ \\
\hline 
B & Warm & $0.35_{-0.04}^{+0.05}$ & $170_{-20}^{+30}$ & $6.6_{-1.9}^{+4.1}\times10^{18}$ \\
  & Hot & $\gtrsim0.25$ & $220_{-20}^{+20}$ & $1.6_{-0.9}^{+1.8}\times10^{18}$ \\
\enddata
\end{deluxetable}

\section{Discussion} \label{sec:discussion}
\subsection{Comparison with the Previous Observations}
Here we compare the distributions of the NH$_3$ emission and physical parameters derived in Section \ref{sec:analysis} with the previous molecular line observations with ALMA for each source A and source B.
\subsubsection{Source A}\label{subsubsec:comparison_sourceA}
The circumbinary structure of source A has been investigated in detail by several previous studies \citep[e.g.,][]{Oya2016, vantHoff2020}. \citet{Oya2016} found that the distributions of CH$_3$OH and HCOOCH$_3$ emission are concentrated in the central region ($\lesssim100$\,au radius) with a clear velocity gradient while the OCS emission is more extended ($\gtrsim200$\,au radius). The spatial extent, as well as the velocity structure, of the NH$_3$ emission in the circumbinary region in the present study is similar to the COM emission rather than the OCS emission, although the spatial extent also depends on the excitation level of each transition (Figure \ref{fig:sourceA_mom0}). This match in the spatial extent suggests a common origin between COMs and NH$_3$ (see Section \ref{subsec:origin_of_ammonia_emission}). Indeed, the NH$_3$ gas temperatures derived by the excitation analysis ($\sim100$\,K at the offset position and $\sim100$--300\,K at the protostar A1 position) are comparable to the value derived by the spatially-resolved line ratio and/or rotation diagram analysis of multiple H$_2$CS lines \citep[$\sim70$--200\,K at circumbinary to disk scales;][]{Oya2016, vantHoff2020}.    

At a smaller scale, recent high-resolution observations directly resolve the binary protostars A1 and A2 within source A in both dust continuum and molecular line emission at $\lesssim10$\,au scale \citep{Maureira2020, Oya2020, Maureira2022}. \citet{Oya2020} found a sudden temperature increase ($\gtrsim200$\,K) inside source A by the line ratio and rotation diagram analysis of the high-resolution ALMA data of the H$_2$CS lines. \citet{Maureira2022} also found a local temperature increase ($\gtrsim100$--300\,K, depending on the assumption of the spectral index) at the location near the protostar A1 (``p2'' in their work, $\sim0\farcs1$  southwestern side of A1) by the multi-band continuum observations with ALMA. The spatially resolved map of an HNCO line also peaks around p2 \citep{Maureira2020, Maureira2022}. The spatial peaks of the observed high excitation NH$_3$ lines (Figure \ref{fig:sourceA_chanmap_highJ}), although unresolved, are coincident with the locations of this ``hot spot'', suggesting that the high excitation transitions of NH$_3$ selectively trace this hot region. The high NH$_3$ gas temperature ($\sim310$\,K) and the small beam filling factor ($\sim0.34$, corresponding to a source size of $\sim0\farcs3$) of the hot component derived by the excitation analysis are also consistent with the temperature and size of this hot spot. The peak velocity of the high-$J$ NH$_3$ lines ($\sim5$--6\,km s$^{-1}$; see Figure \ref{fig:spectrum}) is also similar to the peak velocities of the spatially resolved HNCO and NH$_2$CHO line data ($\sim5.0$ km s$^{-1}$) that have emission peaks at this hot spot \citep{Maureira2022}, further supporting this interpretation. 

\subsubsection{Source B}\label{subsubsec:sourceB}
The physical and chemical structure of source B at $\sim100$\,au scale are studied in detail with ALMA by \citet{Oya2018}. They found a similar chemical differentiation to the source A case where the OCS and H$_2$CS emission are extended while the emission of COMs such as CH$_3$OH and HCOOCH$_3$ shows more compact distributions. Similar to the source A case, the emission distributions of the low excitation NH$_3$ lines are consistent with those of the COM emission, where the depression at the center is seen in both NH$_3$ and COM emission (see Figure \ref{fig:comparison_stacked_highE_lowEu} in the present work and Figure 7 in \citealt{Oya2018}). 

One of the notable differences from the COM emission observed at (sub-)mm wavelengths is the presence of the compact emission of the high-excitation lines and the lack of the absorption feature at the redshifted velocities, i.e., the inverse P-Cygni profile. The absorption feature at the redshifted velocities has ubiquitously been seen by different molecular lines in previous ALMA observations \citep{Pineda2012, Jorgensen2012, Zapata2013, Oya2018}, which is due to the line absorption against the bright continuum emission as a background, and indicates the infalling motion along the line of sight \citep{Myers1996, DiFrancesco2001}. In contrast, as briefly mentioned in Section \ref{subsec:distributions}, almost all NH$_3$ lines do not show any absorption (i.e., negative) features while some of the lines hinting a blue-skewed profile (i.e., weaker redshifted emission but without negatives) after line stacking analysis
\footnote{Note that to produce the blue-skewed profile, the line emission need to have some opacity. Usually non-metastable lines are optically thin, but judging from their similar peak brightness temperatures to those of optically thick metastable lines (see, e.g., Figure \ref{fig:spectrum}), they are likely (at least marginally) optically thick in source B.}. 
This difference is likely due to the combination of the two effects, (1) the weaker (or compact) background continuum emission in the present VLA observations, and (2) the NH$_3$ line emission
traces the warmer component than that traced by (sub-)mm observations with ALMA. This is fully consistent with the fact that the dust continuum emission is more compact and partially optically thin at cm wavelengths \citep{Hernandez-Gomez2019}: the extent of the continuum emission decreases as the observing wavelength increases at cm wavelengths ($\lesssim50$\,GHz) while it is constant in the (sub-)mm wavelengths ($\gtrsim200$\,GHz), meaning that the optically-thin outer region of the continuum allows the line emission to appear there at cm wavelengths. The high gas temperature of the hot component ($\sim220$\,K) derived by the excitation analysis is also consistent with the high-resolution brightness temperature map of the VLA continuum observations at $0\farcs1$--$0\farcs2$ scale \citep{Hernandez-Gomez2019, Zamponi2024}. These facts mean that the VLA observations uniquely trace the inner, hot molecular gas hidden by the optically thick dust emission in the (sub-)mm wavelengths. However, the peak brightness temperature of the high-resolution ($\lesssim0\farcs1$) dust continuum emission at 30--40\,GHz reaches $\sim700$--900\,K \citep{Hernandez-Gomez2019}, which is a proxy of temperature in the innermost region and is far higher than the NH$_3$ gas temperature derived here. This may mean that the hot gas in the most central region is still hidden by the optically thick dust even at cm wavelengths.



\subsection{Origin of the \texorpdfstring{NH$_3$}{} Emission}\label{subsec:origin_of_ammonia_emission}
For both source A and source B, the observed NH$_3$ emission distributes around the protostars up to $\sim50$--100\,au scales. The distributions are similar to those of the COM emission as observed with ALMA \citep[e.g.,][]{Oya2016, Oya2018}. This indicates that NH$_3$ traces so-called ``hot corino'' chemistry, where icy molecules are released into the gas phase via heating and subsequent desorption from the dust grain surfaces. This is similar to the case of NGC1333~IRAS4A, where \citet{DeSimone2022} and \citet{Yamato2022} have detected warm ($\gtrsim100$\,K) NH$_3$ gas at $\sim1\arcsec$ (or $\sim300$\,au) resolution. 
The binding energy of NH$_3$ has been estimated to be $\sim3000$--5000\,K, which is similar to that of CH$_3$OH and other COMs \citep[][and references therein]{Minissale2022}.
Experimental studies suggest that the desorption of NH$_3$ on the interstellar ice could be somewhat complex, where the pure NH$_3$ ice will desorb at $\sim100$\,K and then the NH$_3$ ice entrapped within the water ice matrix will co-desorb at $\sim160$\,K \citep{Martin-Domenech2014, Kakkenpara2024}. The gas temperatures derived by the multi-line analysis are comparable to or higher than these temperatures, and we thus suggest that the observed NH$_3$ emission has an ice origin.

Recent high-resolution observations have revealed the localized emission of COMs and/or temperature enhancements at the disk-envelope interfaces, sometimes associated with streamers, suggesting that the shock heating could be a common mechanism to release icy molecules into gas phase \citep[e.g.,][]{OKoda2022, Kido2023}. As mentioned in Section \ref{subsubsec:comparison_sourceA}, a localized temperature enhancement and associated COM emission at a slightly offset position from the protostar A1 has been observed in source A \citep{Oya2020, Maureira2022}, suggesting shock heating origin. The detection of localized emission of high excitation NH$_3$ emission at the same position and velocity in the present study suggests that the high excitation NH$_3$ lines trace the shocks. Indeed, theoretical computation on the release of icy molecules by shock heating suggest that the weak shock (a shock velocity of a few km s$^{-1}$) can release the molecules that have a binding energy similar to NH$_3$ at preshock gas densities of $\gtrsim10^7$ cm$^{-1}$, depending on the dust grain size \citep{Aota2015, Miura2017, vanGelder2021}. The line widths of the high excitation NH$_3$ lines (a few km s$^{-1}$) are also consistent with this prediction. 

In contrast, the NH$_3$ emission in source B does not show any complex spatiokinematic structures and peaks at the continuum peak (i.e., protostellar position). One of the possible origins of the high excitation NH$_3$ lines in source B is thus the accretion heating caused by gravitational instability \citep{Zamponi2021}, as also indicated by the high brightness temperature at the peak of the continuum emission (reaching $\sim900$\,K at 33\,GHz and $\sim0\farcs1$ resolution; \citealt{Hernandez-Gomez2019}), rather than localized shocks. The smaller temperature difference in the two-component model fit, as well as a reasonable fit with a single-temperature model, also suggest a rather smooth physical structure and no physically distinctive components.

The hot components of the NH$_3$ emission, particularly in source A with a larger temperature difference from the warm component, potentially suggest an additional supply of NH$_3$ into the gas phase. \citet{Maureira2022} argued that the hot spots in source A are explained by localized shocks and derived a postshock temperature of $\sim400$\,K and a shock width of $\sim5$\,au using a simple strong adiabatic shock prescription. Using this rough estimate of shock width and the NH$_3$ column density derived in Section \ref{subsec:fit_protostar}, we can roughly estimate the NH$_3$ gas density enhancement at the shocks. The NH$_3$ gas density at the shocks is $5\times10^4$\,cm$^{-3}$, while the disk-averaged NH$_3$ gas density is ($3$--$5)\times10^3$\,cm$^{-3}$, where we assumed a circumbinary disk radius of 50--100\,au and an inclination angle of $60\arcdeg$. This factor of $10$--$20$ increase in density is larger than the maximum expected 
ratio of the postshock to preshock gas density
of $\sim6$ \citep{Maureira2022}, perhaps implying an additional release of NH$_3$ from the solid phase at the shocks. We note that, however, the rough estimates above are sensitive to a number of assumptions, including shock pitch angle, shock speed, and dust opacity. We need more sophisticated modeling of the shocks, or an estimates on the abundance ratio change at the shocks, to better infer the behavior of NH$_3$ at the shocks.  

This potential additional release of NH$_3$ can be attributed to, for example, the desorption of ammonium salt. Ammonium salt is found to be abundant in the comet 67P/Churyumov-Gerasimenko \citep{Poch2020, Altwegg2020} with a potential major form of ammonium hydrosulfide (NH$_4$SH; \citealt{Altwegg2022}). The presence of NH$_4$SH is also suggested in the ISM by the ice observations with JWST, which may account for $\sim10$\% of the total nitrogen (and sulfur) budget \citep{Slavicinska2024}. Ammonium salt is expected to readily dissociate into NH$_3$ and the acid counterpart after the sublimation, and may thus provide us with a chance to observe as NH$_3$ line emission. The sublimation temperatures of ammonium salt are ranging from $\sim150$\,K to $\sim300$\,K depending on the acid counterparts \citep{Bossa2008, Danger2011, Bergner2016, Vitorino2024, Slavicinska2024}, which is comparable to the gas temperature of the hot components in both source A and source B. Future observations of the acid counterparts may be able to identify the desorption of ammonium salt.



\subsection{Molecular Abundance Ratios}

\begin{deluxetable*}{lcccc}
\tablecaption{Comparison of NH$_3$ Abundance Ratios \label{tab:abundance_ratios}}
\tablewidth{0pt}
\tablehead{
    \colhead{} & 
    \colhead{NH$_3$ / H$_2$O (\%)} & 
    \colhead{NH$_3$ / CH$_3$OH} & 
    \colhead{NH$_3$ / (CH$_3$CN + CH$_3$NC)} & 
    \colhead{References}
}
\startdata
\textit{Molecular Clouds} \\
\quad NIR38   & $4.4_{2.1}^{10.6}$\tablenotemark{$^{a}$} & 0.49\tablenotemark{$^{b}$} & \nodata & (1), (2) \\
\quad J110621 & $5.5_{2.5}^{8.6}$\tablenotemark{$^{a}$} & 1.3\tablenotemark{$^{b}$} & \nodata & (1), (2) \\
\hline
\textit{Low-mass Protostars (ice)} \\
\quad Summary in \citet{Boogert2015} & $6_4^8$\tablenotemark{$^{a}$} & 1\tablenotemark{$^b$} & \nodata & (3) \\
\quad Ced 110 IRS4A\tablenotemark{$^{c}$} & $6.4$ & 3.0\tablenotemark{$^{b}$} & \nodata & (2) \\
\quad IRAS2A & $5.00$ & 1.2\tablenotemark{$^{b}$} & \nodata & (4) \\
\hline
\textit{Low-mass Protostars (warm gas)} \\
\quad IRAS 16293-2422 & $\sim1$--$10$ & $\sim0.01$--$0.1$ & $\sim60$ & (5) \\
\quad NGC1333 IRAS4A\tablenotemark{$^{c}$}  & \nodata & 0.015–0.5 & \nodata & (6) \\
\hline
\textit{Solar System Comets} \\
\quad 67P/Churyumov-Gerasimenko  & $0.67_{-0.20}^{+0.20}$ & $3.2$\tablenotemark{$^{b}$} & $\sim110$ & (7), (8) \\
\enddata
\tablecomments{$^a$Median and lower and upper quartile values. $^b$Calculated from the literature (median) values of NH$_3$ and CH$_3$OH abundances relative to H$_2$O. $^c$Values for one of the binary sources, both of which yield similar values.}
\tablerefs{
    (1) \citet{McClure2023}; (2) \citet{Rocha2025}; (3) \citet{Boogert2015}; (4) \citet{Rayalacheruvu2025}; (5) This work; (6) \citet{DeSimone2022}; (7) \citet{Rubin2019}; (8) \citet{Drozdovskaya2019}
}
\end{deluxetable*}

Here we discuss the chemistry related to NH$_3$ based on the column density ratios of major icy molecules, where column densities of molecules other than NH$_3$ are taken from literature. \revise{Table \ref{tab:abundance_ratios} summarizes the comparison of NH$_3$ abundance ratios among different evolutionary stages discussed below.}

\citet{Persson2013} detected H$_2^{18}$O (692 GHz and 203 GHz) emission with ALMA and the Submillimeter Array (SMA), and derived an excitation temperature of $124\pm12$ K and an H$_2$O column density of $5.3 \times 10^{20}$ cm$^{-2}$ toward source A, assuming a $^{16}$O/$^{18}$O ratio of 560. Comparing with the NH$_3$ column densities at the protostar position of source A, the NH$_3$/H$_2$O ratios are on the order of 10$^{-2}$--10$^{-1}$ \revise{(or 1--10\%)}, broadly consistent with the ice abundance in low-mass protostars \revise{\citep{Oberg2011_icelegacy, Boogert2015, Rocha2025, Rayalacheruvu2025}} and molecular clouds \citep{McClure2023}. 

The column density of CH$_3$OH, another abundant icy molecules, has been estimated in both source A and source B by several studies at the offset positions same as used here to be $\sim1\times10^{19}$ cm$^{-2}$ for source A \citep{Manigand2020} and $\sim1$--$4\times10^{19}$ cm$^{-2}$ for source B \citep{Jorgensen2018, Nazari2024} with an assumption of 0\farcs5 source size. The NH$_3$/CH$_3$OH ratios at the offset positions are on the order of 10$^{-2}$--10$^{-1}$, which are lower than the typical ice abundances \revise{in low-mass protostars and molecular clouds} \revise{\citep{Oberg2011_icelegacy, Boogert2015, Rocha2025, McClure2023, Rayalacheruvu2025}}. As both NH$_3$ and CH$_3$OH mainly form on the dust grain surface via the hydrogenation of N atom and CO molecule, respectively \revise{\citep{Watanabe2002, Hidaka2011, Fedoseev2015}}, the low NH$_3$/CH$_3$OH ratio may reflect the dust temperature of the parental core where these molecules form, as suggested by \citet{DeSimone2022}. While CH$_3$OH can form even with the warmer dust temperature up to $\sim25$--$30$\,K, NH$_3$ formation is less efficient at warmer temperature due to the shorter residence time of N atoms \revise{\citep{Fedoseev2015, DeSimone2022}}, resulting in low NH$_3$/CH$_3$OH ratios. Indeed, the dust temperature of the cloud surrounding the IRAS 16293-2422 system derived from \textit{Herschel} data is 16--17 K \citep{Ladjelate2020}, slightly warmer than the typical $\sim10$\,K and similar to the dust temperature in NGC 1333 IRAS 4 region, where similarly low NH$_3$/CH$_3$OH ratios have been obtained toward three protostars \citep{DeSimone2022}. 

\citet{Drozdovskaya2019} performed a comprehensive comparison of the molecular abundance ratios between IRAS 16293-2422 B and comet 67P for each major volatile element. They compare the abundances of nitrogen-bearing molecules relative to CH$_3$CN (plus its isomer CH$_3$NC as they are indistinguishable in mass spectrometer used for abundance estimation in comet 67P), and use an NH$_3$ column density estimate (upper limit) from single-dish observations for deriving NH$_3$/(CH$_3$CN + CH$_3$NC) ratio in source B ($\lesssim1.5\times10^3$). We update this ratio using the NH$_3$ column density at the offset position to be $\approx60$, which better agrees with the value in comet 67P ($\approx110$) within a factor of $\sim2$. \revise{This agreement might indicate that the nitrogen partitioning between different nitrogen-bearing species in protostellar phase is partially inherited to comet 67P, although more exhaustive comparison among various species, which is beyond the scope of this paper, is needed for a firm conclusion.} 

Note that particular attention should be paid to the fact that in the above abundance ratio calculations, we primarily rely on the literature values of column densities mostly based on the (sub-)mm observations \revise{where the high dust optical depth \rrevise{may significantly reduce the intensity of the molecular line emission},} while present observations of NH$_3$ are performed at cm wavelengths. Furthermore, other observational conditions, such as spatial resolution and sensitivity, also differ for each molecule. These variations can introduce significant uncertainty into the aforementioned abundance ratios, and they should therefore be treated as tentative values. More reliable inferences require uniform observations and analytical methodologies.

\section{Summary}\label{sec:summary}
We presented the first spatially-resolved, multi-line observations of NH$_3$ lines with VLA at high resolution of $\sim0\farcs5$ toward the Class 0 source IRAS 16293-2422. This dataset includes 17 inversion transitions of NH$_3$ with a wide range of upper state energies spanning from $\sim23$\,K to $\sim1,580$\,K, allowing us to constrain the physical condition of the NH$_3$ gas at $\sim50$--100\,au scale in detail. Our conclusion is summarized as follows:

\begin{enumerate}
    \item We detected a number of metastable ($J=K$) and non-metastable ($J\neq K$) NH$_3$ inversion transitions toward both source A and source B, including high excitation lines ($E_\mathrm{u} > 1,000$\,K) for the first time in low-mass star-forming regions. While the high excitation lines show compact distributions in the vicinity of protostars, emission of the low excitation lines ($E_\mathrm{u}\lesssim150$\,K) is more extended at $\gtrsim100$\,au scale. 
    \item For source A, the low excitation NH$_3$ emission shows a clear velocity gradient along the northeast-southwest direction, while the high excitation lines show compact distributions and no clear velocity gradients. The distribution and kinematics of the low excitation lines are consistent with those of the COM emission observed with ALMA. The high excitation lines also show stronger redshifted emission than the blueshifted counterpart.
    \item For source B, the low excitation lines exhibit a depression at the central region, which is similar to the COM emission observed with ALMA. In contrast, the high-excitation lines appears as centrally peaked and more compact.
    \item We constrained the rotation temperature and column density of NH$_3$ at the offset positions as in the previous PILS studies. The rotation temperature is higher than $\sim100$\,K, indicating that the NH$_3$ emission traces the warm gas around the protostars. We also constrained them at the protostar position utilizing two-component model, which provide the constraints on the properties of the hot NH$_3$ gas as traced by high-excitation lines. The rotation temperature of the warmer component is $\sim200$--300\,K, suggesting that the high excitation lines selectively trace hot gas in the vicinity of protostars.
    \item For source A, we suggest that the warmer component is originated from the local shock heating, based on the high temperature of $\sim300$\,K and the spatial and velocity correspondence of the high excitation NH$_3$ lines to the hot spot identified by the high-resolution ($\sim10$\,au) ALMA observations. For source B, the origin of the hot NH$_3$ gas could be the accretion heating potentially triggered by the gravitational instability. 
    \item We compute the abundance ratios of NH$_3$ with respect to two abundant icy molecules, H$_2$O and CH$_3$OH. While NH$_3$/H$_2$O ratio is consistent with the ice abundance ratios in protostellar environments and molecular clouds, NH$_3$/CH$_3$OH ratio is lower than the typical ice abundance ratio. This may indicate that the slightly warmer ($\sim16$--17\,K) condition of the parent core during the prestellar phase inhibits the formation of NH$_3$ ice as also suggested in NGC 1333 IRAS 4 region.  
\end{enumerate}

Given the unique capability of observing NH$_3$ inversion transitions, the lower optical depth of the dust thermal emission, and the lower line density, (sub-)centimeter observations are crucial in unveiling the nitrogen content, or, more generally, molecular composition in protostellar environments. The NH$_3$ inversion transitions are accessible only in this wavelength range, while deuterated isotopologues are also available at (sub-)millimeter wavelengths 
\revise{where severe line confusion from COMs and other species prevents their secure identification for line-rich sources. The line confusion is less severe for more evolved sources, such as protoplanetary disks, but sensitivity remains the primary bottleneck for observing these isotopologues.}
The optically thick dust thermal emission can also obscure the line emission at the (sub-)millimeter wavelengths, while at centimeter wavelengths molecular line emission is indeed visible as shown in the present work \citep[see also][]{DeSimone2020}. Moreover, the present work suggests that the line selection is important in disentangling the different emission components, as only the high excitation NH$_3$ lines ($E_\mathrm{u}>1000$\,K) selectively trace the hot gas in the vicinity of the protostars.  

In the near future, the Next Generation Very Large Array (ngVLA) will be a key facility in observing NH$_3$ lines in the innermost region ($\lesssim10$\,au) of the protostellar / protoplanetary disks that is not spatially resolved at the present VLA observations. The present study will be a critical benchmark for the future observations with such upcoming facilities. 


\begin{acknowledgments}
\revise{We thank the anonymous reviewer for their constructive comments that improved the readability of the manuscript.} Y.Y. is financially supported by Grant-in-Aid for the Japan Society for the Promotion of Science (JSPS) Fellows (KAKENHI Grant Number JP23KJ0636) and the RIKEN Special Postdoctoral
Researcher Program (Fellowships). The National Radio Astronomy Observatory and Green Bank Observatory are facilities of the U.S. National Science Foundation operated under cooperative agreement by Associated Universities, Inc.
\end{acknowledgments}





%
\facilities{VLA}

\software{CASA \citep{CASA}, \texttt{bettermoments} \citep{bettermoments}}


\appendix

\section{Hyperfine Spectroscopic Data of the Observed \texorpdfstring{NH$_3$}{} Transitions}

Table \ref{tab:transitions_hfs} list the hyperfine spectroscopic data of the observed NH$_3$ inversion transitions, which is used for the multi-line analysis. Note that the hyperfine spectroscopic data are available up to NH$_3$ (8,6). 

\startlongtable
\begin{deluxetable}{cCCCRC}
\label{tab:transitions_hfs}
\tablecaption{Hyperfine Spectroscopic Data of NH$_3$ Transitions}
\tablehead{\colhead{Transition} & \colhead{$F'$} & \colhead{$F''$} & \colhead{$\nu_0$} & \colhead{$r$} & \colhead{$\delta v$} \\ 
\colhead{$(J, K)$} & \colhead{} & \colhead{} & \colhead{(GHz)} & \colhead{} & \colhead{(km s$^{-1}$)}}
\startdata
(1,1) & 0 & 1 & 23.692963 & 0.1111 & 19.4 \\
 & 2 & 1 & 23.693882 & 0.1389 & 7.8 \\
 & 1 & 1 & 23.694496 & 0.0833 & 0.0 \\
 & 2 & 2 & 23.694496 & 0.4166 & 0.0 \\
 & 1 & 2 & 23.695108 & 0.1389 & -7.8 \\
 & 1 & 0 & 23.696030 & 0.1111 & -19.4 \\
\hline
(2,2) & 1 & 2 & 23.720587 & 0.0500 & 25.9 \\
 & 3 & 2 & 23.721317 & 0.0518 & 16.6 \\
 & 2 & 2 & 23.722631 & 0.2315 & 0.0 \\
 & 1 & 1 & 23.722631 & 0.1500 & 0.0 \\
 & 3 & 3 & 23.722631 & 0.4148 & 0.0 \\
 & 2 & 3 & 23.723946 & 0.0519 & -16.6 \\
 & 2 & 1 & 23.724676 & 0.0500 & -25.8 \\
\hline
(3,3) & 2 & 3 & 23.867828 & 0.0265 & 28.9 \\
 & 4 & 3 & 23.868425 & 0.0268 & 21.4 \\
 & 3 & 3 & 23.870131 & 0.2801 & 0.0 \\
 & 2 & 2 & 23.870131 & 0.2116 & 0.0 \\
 & 4 & 4 & 23.870131 & 0.4018 & 0.0 \\
 & 3 & 4 & 23.871834 & 0.0268 & -21.4 \\
 & 3 & 2 & 23.872431 & 0.0265 & -28.9 \\
\hline
(4,4) & 3 & 4 & 24.136961 & 0.0162 & 30.5 \\
 & 5 & 4 & 24.137463 & 0.0163 & 24.3 \\
 & 4 & 4 & 24.139418 & 0.3009 & 0.0 \\
 & 3 & 3 & 24.139418 & 0.2430 & 0.0 \\
 & 5 & 5 & 24.139418 & 0.3911 & 0.0 \\
 & 4 & 5 & 24.141371 & 0.0163 & -24.3 \\
 & 4 & 3 & 24.141872 & 0.0162 & -30.5 \\
\hline
(5,1) & 5 & 4 & 19.836824 & 0.0109 & 23.0 \\
 & 5 & 6 & 19.837082 & 0.0109 & 19.1 \\
 & 4 & 4 & 19.838346 & 0.2618 & 0.0 \\
 & 5 & 5 & 19.838346 & 0.3115 & 0.0 \\
 & 6 & 6 & 19.838346 & 0.3830 & 0.0 \\
 & 6 & 5 & 19.839611 & 0.0109 & -19.1 \\
 & 4 & 5 & 19.839869 & 0.0109 & -23.0 \\
\hline
(5,4) & 4 & 5 & 22.652000 & 0.0109 & 13.5 \\
 & 6 & 5 & 22.652174 & 0.0109 & 11.2 \\
 & 5 & 5 & 22.653023 & 0.3115 & 0.0 \\
 & 4 & 4 & 22.653023 & 0.2618 & 0.0 \\
 & 6 & 6 & 22.653023 & 0.3830 & 0.0 \\
 & 5 & 6 & 22.653870 & 0.0109 & -11.2 \\
 & 5 & 4 & 22.654043 & 0.0109 & -13.5 \\
\hline
(5,5) & 4 & 5 & 24.530424 & 0.0109 & 31.3 \\
 & 6 & 5 & 24.530859 & 0.0109 & 26.0 \\
 & 5 & 5 & 24.532986 & 0.3115 & 0.0 \\ 
 & 4 & 4 & 24.532986 & 0.2618 & 0.0 \\
 & 6 & 6 & 24.532986 & 0.3830 & 0.0 \\
 & 5 & 6 & 24.535112 & 0.0109 & -25.9 \\
 & 5 & 4 & 24.535545 & 0.0109 & -31.2 \\
\hline
(6,2) & 6 & 5 & 18.883511 & 0.0078 & 18.8 \\
 & 6 & 7 & 18.883682 & 0.0078 & 16.1 \\
 & 5 & 5 & 18.884695 & 0.2742 & 0.0 \\
 & 6 & 6 & 18.884695 & 0.3177 & 0.0 \\
 & 7 & 7 & 18.884695 & 0.3768 & 0.0 \\
 & 7 & 6 & 18.885708 & 0.0078 & -16.1 \\
 & 5 & 6 & 18.885880 & 0.0078 & -18.8 \\
\hline
(6,3) & 6 & 5 & 19.756945 & 0.0078 & 9.0 \\
 & 6 & 7 & 19.757030 & 0.0078 & 7.7 \\
 & 5 & 5 & 19.757538 & 0.2742 & 0.0 \\
 & 6 & 6 & 19.757538 & 0.3177 & 0.0 \\
 & 7 & 7 & 19.757538 & 0.3768 & 0.0 \\
 & 7 & 6 & 19.758045 & 0.0078 & -7.7 \\
 & 5 & 6 & 19.758131 & 0.0078 & -9.0 \\
\hline
(6,6) & 5 & 6 & 25.053389 & 0.0078 & 31.5 \\
 & 7 & 6 & 25.053770 & 0.0078 & 27.0 \\
 & 6 & 6 & 25.056025 & 0.3176 & 0.0 \\
 & 5 & 5 & 25.056025 & 0.2742 & 0.0 \\
 & 7 & 7 & 25.056025 & 0.3767 & 0.0 \\
 & 6 & 7 & 25.058281 & 0.0079 & -27.0 \\
 & 6 & 5 & 25.058661 & 0.0078 & -31.5 \\
\hline
(7,7) & 6 & 7 & 25.712488 & 0.0059 & 31.4 \\
 & 8 & 7 & 25.712828 & 0.0059 & 27.5 \\
 & 7 & 7 & 25.715181 & 0.3215 & 0.0 \\
 & 6 & 6 & 25.715181 & 0.2830 & 0.0 \\
 & 8 & 8 & 25.715181 & 0.3719 & 0.0 \\
 & 7 & 8 & 25.717537 & 0.0059 & -27.5 \\
 & 7 & 6 & 25.717876 & 0.0059 & -31.4 \\
\hline
(8,6) & 7 & 8 & 20.718407 & 0.0046 & 11.8 \\
 & 9 & 8 & 20.718498 & 0.0046 & 10.5 \\
 & 8 & 8 & 20.719221 & 0.3242 & 0.0 \\
 & 7 & 7 & 20.719221 & 0.2895 & 0.0 \\
 & 9 & 9 & 20.719221 & 0.3679 & 0.0 \\
 & 8 & 9 & 20.719944 & 0.0046 & -10.5 \\
 & 8 & 7 & 20.720036 & 0.0046 & -11.8 \\
\enddata
\tablecomments{\textrm{The spectroscopic data are taken from the Jet Propulsion Laboratory (JPL) database \citep{JPL}. The original data are presented in \citet{Yu2010}.}}
\end{deluxetable}




\section{Channel Maps}\label{appendix:channel_maps}
\begin{figure*}
\plotone{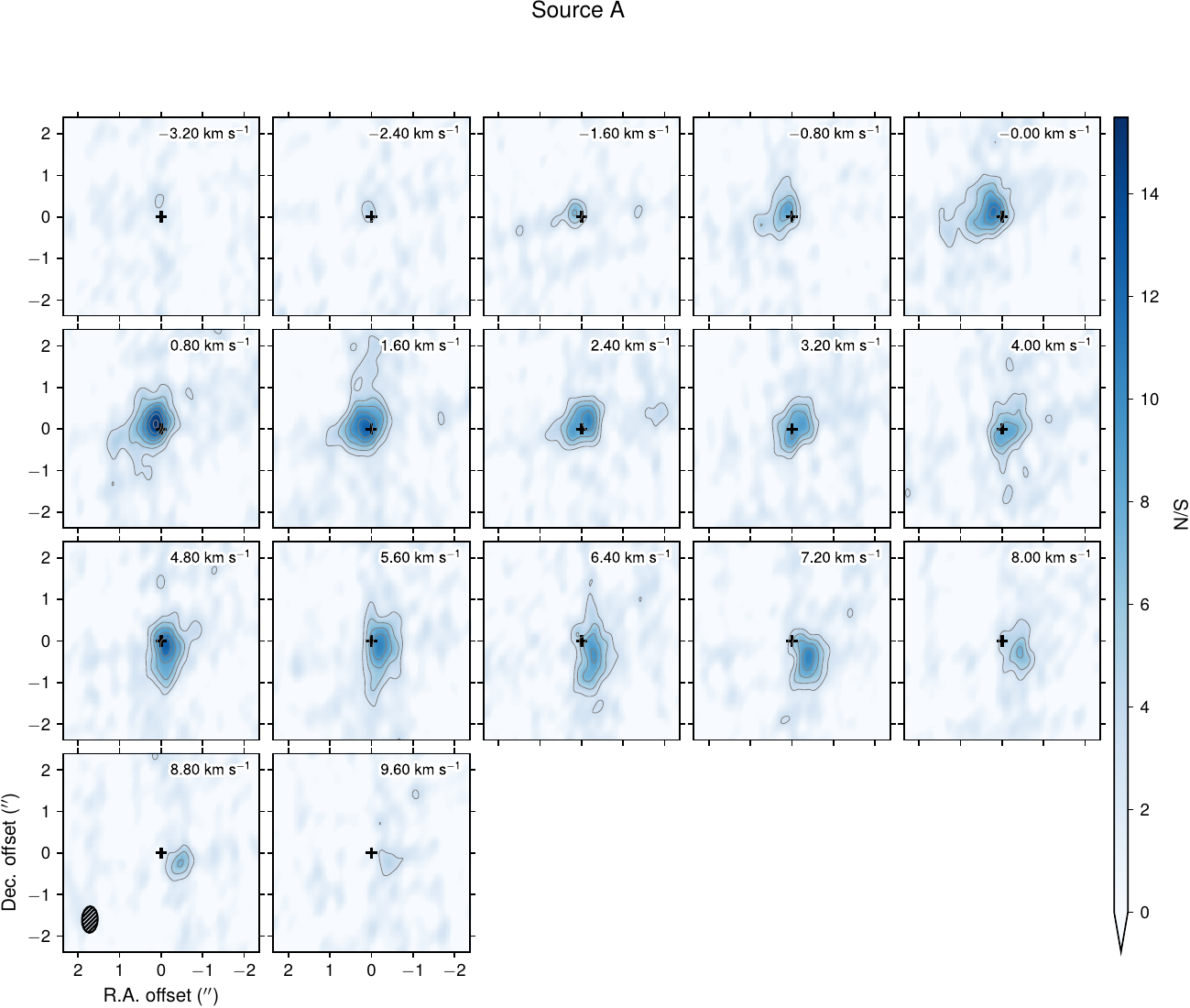}
\caption{Stacked channel maps of the NH$_3$ (2,2) and (3,3) emission in source A shown in units of S/N. Contours start at S/N $=3$ and increase in steps of 2.}
\label{fig:sourceA_chanmap_lowJ}
\end{figure*}

\begin{figure*}
\plotone{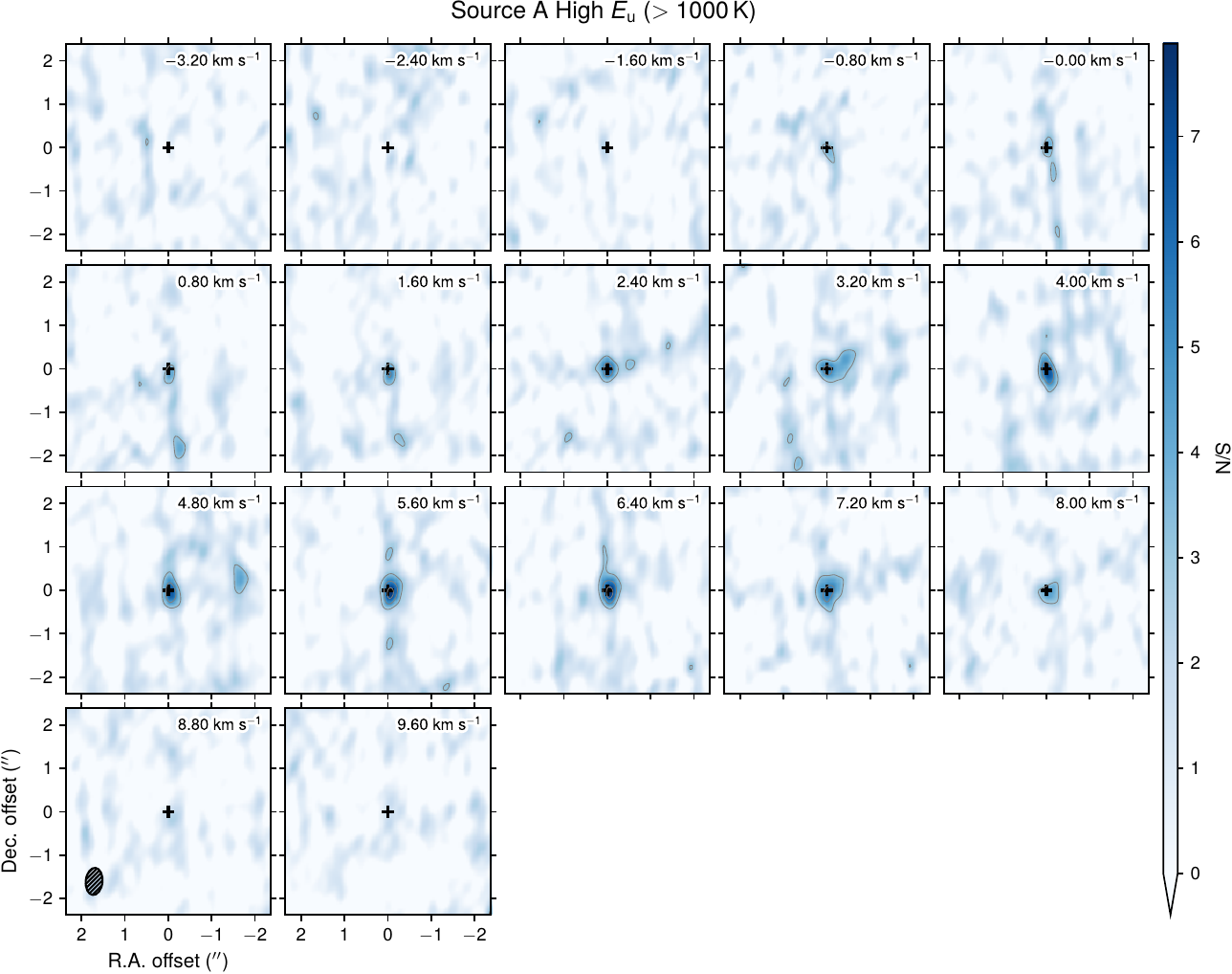}
\caption{Stacked channel maps of transitions with $E_\mathrm{u} > 1000$\,K in source A shown in units of S/N. Contours start at S/N $=3$ and increase in steps of 2.}
\label{fig:sourceA_chanmap_highJ}
\end{figure*}

\begin{figure*}
\plotone{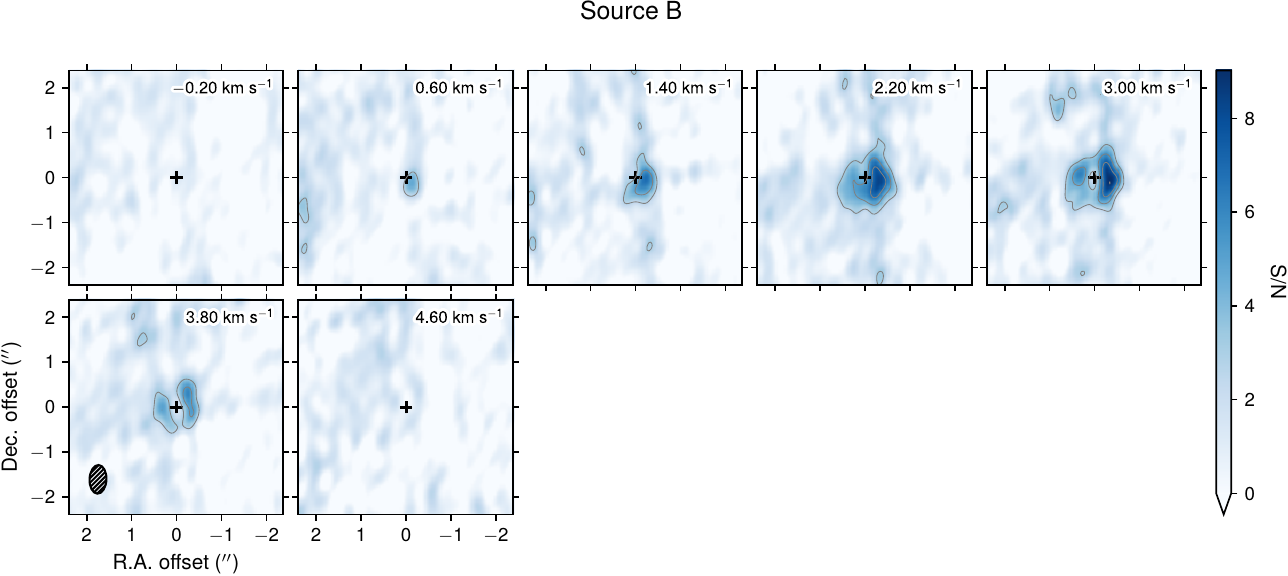}
\caption{Same as Figure \ref{fig:sourceA_chanmap_lowJ}, but for source B. }
\label{fig:sourceB_chanmap_lowJ}
\end{figure*}

\begin{figure*}
\plotone{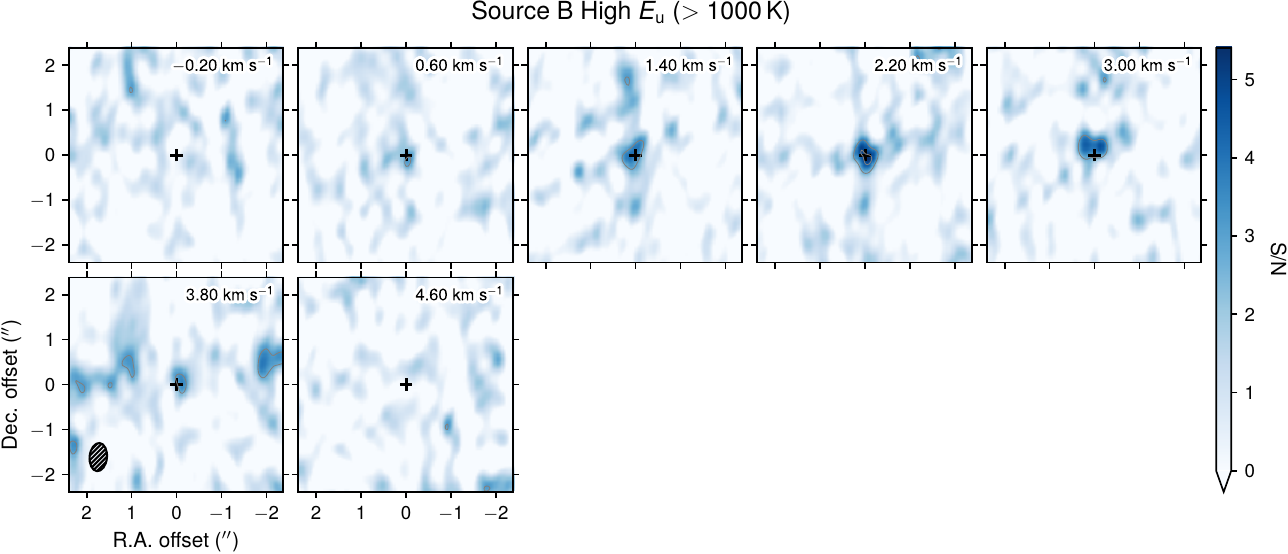}
\caption{Same as Figure \ref{fig:sourceA_chanmap_highJ}, but for source B. }
\label{fig:sourceB_chanmap_highJ}
\end{figure*}

Figures \ref{fig:sourceA_chanmap_lowJ} and \ref{fig:sourceA_chanmap_highJ} show the stacked channel maps of low-$E_\mathrm{u}$ ($<150$\,K, but excluding NH$_3$ (1,1) due to the heavy blending between hyperfine components) and high-$E_\mathrm{u}$ ($>1,000$\,K) NH$_3$ lines in source A, respectively. Similarly, stacked channel maps for source B are shown in Figures \ref{fig:sourceB_chanmap_lowJ} and \ref{fig:sourceB_chanmap_highJ}.

\section{Single Component Fit at the Protostar Position of Source B}\label{appendix:peak_single_fit_sourceB}
\begin{figure}
\plotone{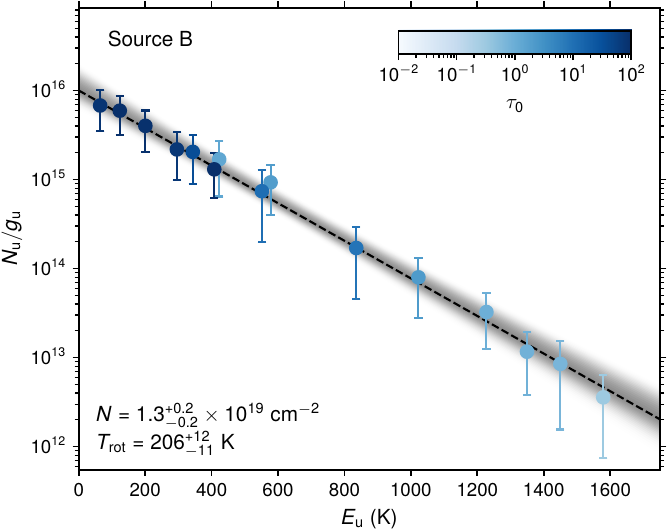}
\caption{Result of the single-component model fit to the velocity-integrated intensities at the protostar positions of source B in the form of rotation diagram. Random drawn from the posterior are shown in gray and the best-fit model that maximize the likelihood is plotted by the black dashed curve. The color represents the maximum optical depth among the two components for each inversion transition.}
\label{fig:fit_peak_single_B}
\end{figure}

Figure \ref{fig:fit_peak_single_B} shows the result of the single-component model fit to the velocity-integrated intensity at the protostar position of source B. We derived the constraints on parameters as $\eta=0.30_{-0.02}^{+0.01}$, $T_\mathrm{rot}=210_{-10}^{+10}$\,K, and $N=1.3_{-0.2}^{+0.2}\times10^{19}$\,cm$^{-2}$.

\bibliography{reference}{}
\bibliographystyle{aasjournalv7}



\end{document}